\documentclass[aps,prl,reprint,superscriptaddress]{revtex4-2}

\usepackage{graphicx}
\usepackage{siunitx}
\usepackage{amsmath}
\usepackage{amssymb}
\usepackage{hyperref}
\usepackage{import} 
\usepackage{caption}
\usepackage[utf8]{inputenc}
\usepackage{booktabs}
\usepackage{wrapfig}
\usepackage{sansmath}

\begin{document}

\preprint{APS/123-QED}

\captionsetup[figure]{labelfont={bf},name={Figure},labelsep=period,justification=raggedright}

\title{
Self-organization of cavity solitons in Brillouin–Kerr ring resonators
}

\author{Corentin Simon}
\affiliation{OPERA-photonics, Université libre de Bruxelles (U.L.B.), 50~Avenue F. D. Roosevelt, CP 194/5, B-1050 Brussels, Belgium}
\author{Matteo Conforti}
\affiliation{University of Lille, CNRS, UMR 8523-PhLAM Physique des Lasers, Atomes et Molécules, F-59000, Lille, France}
\author{Loïc Van Bellinghen}
\affiliation{OPERA-photonics, Université libre de Bruxelles (U.L.B.), 50~Avenue F. D. Roosevelt, CP 194/5, B-1050 Brussels, Belgium}
\author{Maxime Fournier}
\affiliation{OPERA-photonics, Université libre de Bruxelles (U.L.B.), 50~Avenue F. D. Roosevelt, CP 194/5, B-1050 Brussels, Belgium}
\author{Arnaud Mussot}
\affiliation{University of Lille, CNRS, UMR 8523-PhLAM Physique des Lasers, Atomes et Molécules, F-59000, Lille, France}
\author{François Leo}
\affiliation{OPERA-photonics, Université libre de Bruxelles (U.L.B.), 50~Avenue F. D. Roosevelt, CP 194/5, B-1050 Brussels, Belgium}
\author{Simon-Pierre Gorza}
\affiliation{OPERA-photonics, Université libre de Bruxelles (U.L.B.), 50~Avenue F. D. Roosevelt, CP 194/5, B-1050 Brussels, Belgium}

\date{5.~March 2026}

\begin{abstract}
We report on the interaction between stimulated Brillouin scattering and temporal cavity solitons (CSs) in doubly resonant ring resonators. 
Our experiments are performed in coherently driven passive optical-fibre resonators. We demonstrate that the interplay between four-wave mixing and cascade Brillouin lasing spontaneously generates patterns of CSs on a temporal grid at twice the Brillouin-shift. These patterns are shown to be highly stable owing to a long-range locking mechanism mediated by the acoustic oscillation generated by the solitons. We introduce a unified mean-field model of the cavity to describe the dynamics between the coupled forward and backward waves under coherent driving. This model reproduces very well the experiments and explains the paracrystalline structures of the soliton pattern. Our findings significantly advance the understanding of hybrid Brillouin-Kerr optical frequency combs.
\end{abstract}

\maketitle


\textit{Introduction---}Temporal dissipative solitons are particle-like excitations propagating endlessly in optical nonlinear resonators
\,\cite{grelu_dissipative_2012,lugiato_spatial_1987, leoTemporalCavitySolitons2010, herrTemporalSolitonsOptical2014, englebert_parametrically_2021}. 

Among them, temporal cavity solitons (CS)
emerge in coherently driven Kerr resonators, naturally enforcing phase locking between the pulses and the external driving field.  

Over the past two decades, they have attracted considerable attention due to their broad impact across photonics and precision‑measurement technologies\,\cite{fortier20YearsDevelopments2019, marin-palomo_microresonator-based_2017, liu_photonic_2020, newman_architecture_2019, riemensberger_massively_2020}.

In addition to the Kerr‑induced parametric interaction that underpins CSs generation, stimulated Brillouin scattering (SBS) is another prominent nonlinear effect that manifests in optical waveguides\,\cite{kobyakovStimulatedBrillouinScattering2010,eggleton_brillouin_2019}. 
SBS, which couples optical fields to an acoustic wave, provides a narrowband optical gain that enables ultra‑coherent lasing\,\cite{debutLinewidthNarrowingBrillouin2000, gengHighlyStableLownoise2006, liCharacterizationHighCoherence2012, gundavarapuSubhertzFundamentalLinewidth2019, loh_ultra-narrow_2019}, thereby supporting a wide range of applications\,\cite{zarinetchiStimulatedBrillouinFiberoptic1991, liMicroresonatorBrillouinGyroscope2017, murrayDistributedBrillouinFiber2022, loh_operation_2020, liMicrowaveSynthesizerUsing2013, heffernan_brillouin_2024}.

Brillouin scattering has recently attracted renewed attention due to the rich nonlinear dynamics that stem from the interplay between Brillouin lasing and four‑wave mixing, particularly in the context of optical frequency‑comb generation. Non‑reciprocal fiber ring resonators have been explored under both single\,\cite{tang_stable_2011} and multi‑wavelength excitation\,\cite{li_optical_2017, lucasDynamicInterplayKerr2023, deroh_dispersion_2023}, revealing soliton formation with anomalous dispersion\,\cite{huangTemporalSolitonOptical2019}. Brillouin–Kerr frequency combs have also been realized in high‑Q fiber Fabry–P\'{e}rot cavities\,\cite{brajeBrillouinEnhancedHyperparametricGeneration2009, buttner_phase-locking_2014, jia_photonic_2020, bunelBrillouininducedKerrFrequency2025}.
Internally pumped counter‑propagating cavity solitons enabled by Brillouin lasing in microdisks have demonstrated thermal self‑stabilization of the pump and low-noise operation\,\cite{baiBrillouinKerrSolitonFrequency2021}, as well as deterministic single‑soliton states generation\,\cite{do_self-stabilized_2021}. 
Besides, multi‑soliton Kerr combs propagating in opposite directions under a single external pump have been reported in silica microspheres\,\cite{zhangSolitonMicrocombsMultiplexing2023}. 

Investigations of the interaction between cavity solitons and Brillouin scattering have thus far been limited either to non‑reciprocal cavities, i.e., to non‑resonant pumping, or to short resonators whose free‑spectral range exceeds the Brillouin‑gain bandwidth. 
Earlier works on CW-pumped fiber cavities indeed suppressed Brillouin lasing via optical isolation\,\cite{leoTemporalCavitySolitons2010, jang_ultraweak_2013}, despite the CS pump power threshold can be reached first in low‑loss resonators.
Yet, owing to their robust localized nature, the collective dynamics of multiple cavity solitons are expected to be particularly susceptible to Brillouin scattering. 
The formation of multi‑soliton bound‑state patterns is a universal mechanism observed in coherently driven Kerr cavities\,\,\cite{wang_universal_2017, jang_ultraweak_2013, cole_soliton_2017, weng_heteronuclear_2020}, as well as in other dissipative systems such as mode‑locked lasers\,\cite{amrani_universal_2011, nimmesgern_soliton_2021}. 
Hence, the longitudinal acoustic wave emitted by solitons could provide an efficient long‑range binding mechanism, a possibility yet to be explored.

In this Letter, we investigate cavity‑soliton dynamics in doubly resonant Brillouin laser cavities. We introduce a new set of equations that capture the Brillouin–Kerr interaction in driven ring resonators, simplifying previous models\,\cite{dongUnifiedApproachCascaded2016, korobko_detuning_2020}. 
We compare the model’s predictions with experiments in long-fiber resonators and use them to elucidate the spontaneously emerging robust soliton patterns and the associated Kerr comb.

\textit{Results---}

Our theoretical analysis of the intracavity nonlinear dynamics, together with numerical simulations, relies on the coupled mean‑field equations \,\ref{eq1}-\ref{eq2} provided in the End Matter, which are established for doubly resonant Brillouin–Kerr ring lasers.
These equations describe the round-trip evolution of the slowly varying envelopes of the driven forward ($\psi_F$) and backward ($\psi_B$)  fields. 

To get insight into the interplay between CSs and Brillouin lasing, let us analyze the interaction between a single coherently driven forward CS and the first-order Stokes wave $S_1$ [i.e. $\psi_B=S_1 \exp(i\Omega_b \beta_1 z)$, with $\Omega_b=2\pi\nu_b$, the Brillouin shift frequency, $\beta_1$, the inverse of the group velocity and $z$, the position]. 

We first note that the soliton spectrum ($\sim$\,THz) is much wider than the Brillouin gain bandwidth ($\sim 50\,\si{\mega \hertz}$).
When the resonator free spectral range is smaller than the Brillouin bandwidth, the contribution of the CS to the Brillouin interaction term in the equation of $\psi_F$ is proportional to ${h}_b(z_\mathrm{CS} - z) \exp\left[i\Omega_b \beta_1 (z_\mathrm{CS} - z)+i \phi_\mathrm{CS}\right]$ (as detailed in the End Matter). Here, $h_b$ denotes the spatial Brillouin response, and $z_\mathrm{CS}$, $\phi_\mathrm{CS}$ are the soliton position and phase, respectively. 

\begin{figure}
  \begin{center}
    \includegraphics[width=0.8\columnwidth]{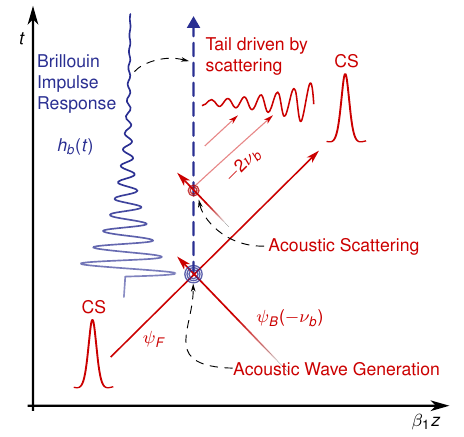}
  \end{center}
  \caption{ 
    Space–time schematic representation of the trailing tail formation behind a CS. 
    The CS locally excites an acoustic wave with the Brillouin impulse response $h_b(t)$, upon which $S_1$ is scattered, resulting in the generation of a decaying trailing wave pinned to the soliton.
  }
  \label{fig1:scheme-int}
\end{figure}

This term acts as an additional coherent drive, composed of two spectral components: one at the frequency of the external driving field and one shifted by $-2\nu_B$.    
The resulting trailing signal, by beating with the forward background field at the driving frequency ($S_0$), generates a stationary pattern in the CS reference frame (see schematic representation in Fig.\ref{fig1:scheme-int}). 
This mechanism creates a long-range oscillating interaction potential with a characteristic timescale of about 100\,ns\,\cite{wang_universal_2017}, enabling mutual interactions among thousands of solitons and potentially facilitating the formation of soliton crystals within the resonator. 

\begin{figure}
  \centering
  \includegraphics{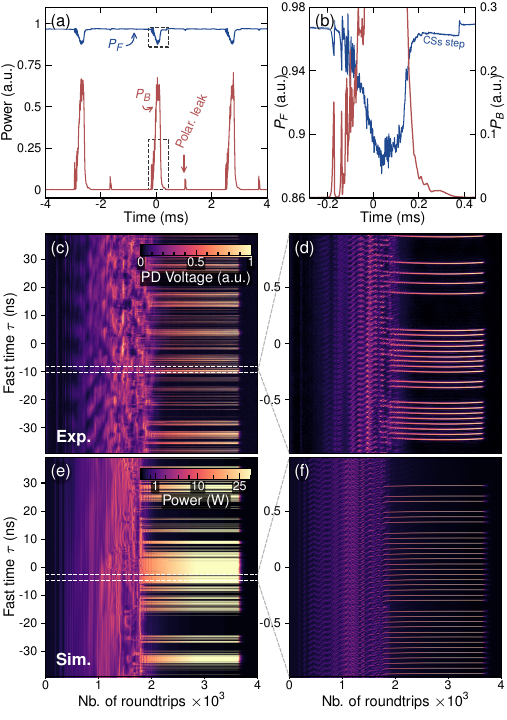}
  \caption{
    (a) Transmitted ($P_F$) and backward ($P_B$) signals when scanning the frequency of a laser over multiple resonances (forward detuning scan rate $\qty{2.3}{\radian / \milli \second}$). The transmission shows the expected tilted nonlinear resonance, followed by a single soliton step. 
    (b) Close-up view of a single resonance. This step co-exists with a weak backward Brillouin lasing signal. 
    (c) Dynamics of the output forward signal, recorded with a high-bandwidth oscilloscope after filtering the CW driving component. 
    The formation dynamics of individual solitons are visible in the zoomed-in view (d).  
    (e,f) The simulation with our mean-field model [Eqs.\ref{eq1}-\ref{eq2}] reproduces very well the whole dynamics ($\qty{2.3}{\radian / \milli \second}$). Raman interaction, which slightly bends CS's trajectory in panel (d), is not included in the simulation for simplicity.
  }
  \label{fig2:scan-dynamics}
\end{figure}

The experiments are performed in a 30.7\,m‑long fiber ring resonator (3.5\% loss, 2.55$\times10^9$ Q-factor) operating in the anomalous‑dispersion regime and driven by a highly coherent continuous‑wave (CW) laser through a 1\% input coupler
(see more details in the End Matter). 
We start by studying the spontaneous formation of CSs in the presence of cascade Brillouin lasing.  
When the intracavity power increases, it eventually crosses the Brillouin laser threshold. Beyond, the first-order backward Stokes wave $S_1$ is generated. 
With a further increase in driving power $P_\mathrm{in}$, or with deeper tuning into resonance, the system crosses successive thresholds that generate odd backward and even forward Stokes orders $S_j$.
This cascade could hinder the spontaneous emergence of CSs during a detuning scan.  
Indeed, the mean-field model without Brillouin shows that a parametric Kerr gain {$P_\mathrm{cav}\gamma L > \Lambda/2$} 
is required for spontaneous CSs formation. Although the Brillouin cascade has little impact on the average loss experienced by CSs, it prevents the build-up of the CW average power $P_\mathrm{cav}$. 
Consequently, we only detect CSs spontaneous generation above $P_\mathrm{in} \gtrsim \qty{500}{\milli \watt}$ while a threshold of $\qty{16}{\milli \watt}$ is predicted without Brillouin lasing.  
The nonlinear resonances of the forward and backward waves are shown in Fig.\,\ref{fig2:scan-dynamics}(a-b) for a driving power of $\qty{600}{\milli \watt}$. We observe a shallow resonance in the average power of the forward signal measured in the through port, despite the resonator being close to critical coupling. Simultaneously, backward lasing occurs, in agreement with the previous analysis of the Brillouin cascade. At higher detunings, i.e., beyond the strong backward lasing peak, the forward output power exhibits the characteristic signature of CS formation, known as the soliton step.
Resonance-scan measurements uncover complex soliton formation dynamics [see Figs. 2(c–d)], with nanosecond-scale transient structures that determine the final CS density as the cascade dissipates.
For convenience, we introduce the fast-time $\tau = -\beta_1 z$, which refers to the time over a single cavity round-trip. 
Zooming in, we see the spontaneous emergence of patterns with occasional stochastic defects.
Strikingly, it consists of individual solitons on a lattice of period $1/2\nu_b$. This period corresponds to the fundamental beating period between even Stokes waves ($2\nu_b= \qty{21.7}{\giga \hertz}$) and arises from the dynamics of the Brillouin cascade as the resonance is scanned (see Supplemental Material Fig.S3(c)). 
Numerical simulations reproduce the resulting soliton patterns with very good agreement [see Figs.\,\ref{fig2:scan-dynamics}(e,f) and Supplemental Material Fig.S3(a-b)], which validate our mean-field model Eqs.\;\ref{eq1}-\ref{eq2}.

\begin{figure}
  \begin{center}
    \includegraphics{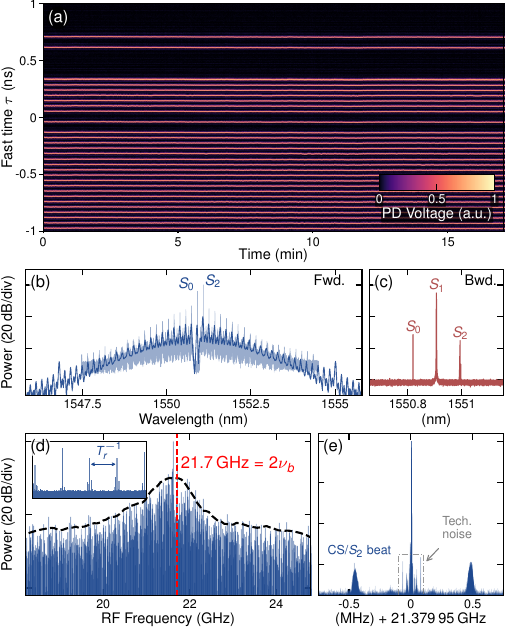}
  \end{center}
  \caption{
    (a) Oscilloscope measurements at the cavity output showing the long-term stability of the soliton pattern ($\delta\approx0.49$). 
    (b) Optical spectrum measured in the through port of the cavity coupler, after pump filtering. Light blue, high resolution (RBW = $\qty{0.16}{\pico \meter}$) and, dark blue, low resolution (RBW = $\qty{30}{\pico \meter}$).   
    (c) Same for the backward signal at high resolution.
    (d) Radiofrequency spectrum corresponding to (b) (different realization). Inset: fine teeth structure of the wide comb.
    (e) Zoom around one RF tooth.
  }
  \label{fig3}
\end{figure}

The long-term stability of the soliton pattern is then studied.
Oscilloscope traces, at a detuning $\delta\approx0.5$, reveal the high stability of the pattern over minutes (Fig.\ref{fig3}a), only limited by the locking range. The backward-recorded spectrum indicates that $S_1$ circulates within the cavity under this stable regime (Fig.\ref{fig3}c). Yet, at higher detunings, it vanishes, and we observe a rapid breaking of the periodicity and the well-known bunching dynamics of CSs stemming from transverse acoustic waves\,\cite{erkintalo_bunching_2015} (not shown). This
highlights the importance of Brillouin-mediated long-range mutual interactions in maintaining solitons on a regular grid.

Perfect cavity-soliton crystallization generates sharp optical spectral lines through multi-soliton coherent interference\,\cite{cole_soliton_2017}.
Stochastic vacancies in such soliton crystals manifest themselves as reduced overall visibility in the optical spectrum. Figure\,\ref{fig3}b shows the spectrum, measured at the cavity output, after filtering the pump-laser frequency. In addition to the intense $S_2$ line, it features narrow peaks separated by a wavelength shift corresponding to $2\nu_b$, in agreement with the temporal pattern. Their visibility, however, decreases away from the pump, suggesting irregular soliton spacing. This is confirmed in the electrical spectrum of the pulse train (Fig.\,\ref{fig3}d). It consists of thousands of beat notes spaced at multiples of the cavity FSR, with a relatively flat structure due to vacancies. It also shows a broad peak at $2\nu_b$, thus lacking the sharp RF beat note expected for a perfect crystal containing only vacancies.
Additionally, a closer inspection reveals broadened sidebands at $\pm \qty{470}{\kilo\hertz}$ from the main RF peaks (see e.g. Fig.\ref{fig3}e).

They originate from the frequency shift between $S_2$ and the closest soliton comb frequency. 
$S_2$, which lies on a cavity mode, indeed does not coincide with a tooth of the soliton comb because of the pump detuning. As a result, the background, modulated by the interference between $S_0$ and $S_2$, drifts in the soliton reference frame. This causes the CSs to oscillate in amplitude at the frequency $\delta/(2\pi T_R)$, with $T_R$ the round-trip time. The beating frequency measured in the RF spectrum gives a detuning $\delta=\qty{0.44\pm 0.01}{}$, in agreement with the detuning estimated from the fit of the optical spectrum [$\delta = \qty{0.45\pm 0.02}{}$]. 
Interestingly, the beating peak width provides a measurement of the detuning stability. We extract a detuning standard deviation of $\sim \qty{10}{\milli \radian}$ for our dedicated stabilization scheme (see Supplemental Material). 

\begin{figure}
  \centering
  \includegraphics{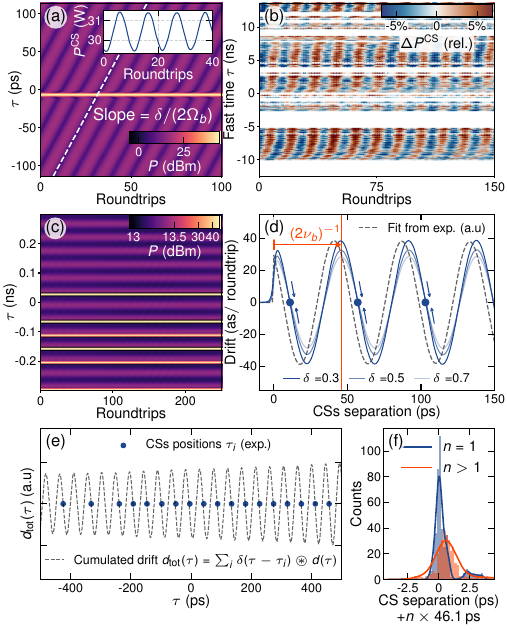}
  \caption{
    (a) Simulation of a CS showing its oscillation induced by the drifting modulated background from the $S_0-S_2$ beating. Inset: soliton peak power variation.
    (b) Experimental measurement of the oscillation of the CSs. Each point of the scatter plot corresponds to a single CS. 
    (c) Simulated intracavity power for spontaneously generated CSs, after numerical filtering of $S_2$. The weak modulation at $2\nu_b$, stationary in the CSs reference frame and responsible for their locking, can be seen. 
    (d) Theoretical soliton drift induced by the interaction between a CS located at $\tau=0$ and $S_1$, computed from Eqs.\,\ref{eq1}-\ref{eq2} (solid lines). Stable equilibrium points are indicated. The dashed line shows our simplified analytical model of the drift, fitted to the experimental CS positions.
    (e) Reconstructed cumulated drift from the fitted model and experimental CSs positions computed from the oscilloscope trace recorded over a round-trip once the pattern is stabilized.
    (f) Distribution of the experimental delay between two consecutive solitons, separated by a single ($n=1$) or multiple ($n>1$) lattice period $1/2\nu_b = \qty{46.1}{\pico \second}$. 
  }
  \label{fig4}
\end{figure}

We exploit the time-domain oscillations to probe the local distance between solitons over the entire cavity round-trip, and gain insight into their locking mechanism.
The numerical simulation reported in Fig.\ref{fig4}a displays the drifting modulated background, along with a single oscillating soliton. For regularly spaced CSs, the phase of the oscillations would be constant or linearly varying with position across the pulse sequence. This is not observed in the experiment. Instead, the direct measurement of the oscillations (Fig.\,\ref{fig4}b) shows that closely spaced CSs are not perfectly equidistant, although their oscillation phases exhibit a certain degree of correlation. In contrast, more distant solitons appear to be randomly positioned. This confirms the irregular spacings in the self‑organization of the solitons, already seen in the optical and RF spectra.
Filtering out, in simulations, the background modulation arising from the $S_0$-$S_2$ beating reveals a residual stationary pattern (Fig.\,\ref{fig4}c). 
This pattern originates from the Brillouin coupling between $S_1$ and each soliton, as predicted, which drives the oscillating tails responsible for its formation.
The comparison between the CSs position and the modulation indicates that this pattern underlies the CS pinning mechanism, in agreement with the observation that the presence of $S_1$ is required for long-term stabilization. The locking is similar to the well-established mechanism attracting CSs to extrema of an external drive amplitude or phase modulation\,\cite{erkintaloPhaseIntensityControl2022}, and provides a many-body interaction between CSs. The theoretical drift that a single soliton induces on another can be determined by projecting the corresponding drive term onto the CS neutral mode \cite{maggipintoCavitySolitonsSemiconductor2000}.
This analysis reveals a one-sided CS-to-CS interaction, leading to stable points arranged in a lattice with a period of $1/2\nu_b$ (Fig.\,\ref{fig4}d). 
For comparison purposes, we fit a similar interaction function [$d(\tau) = \sin(a\tau+b)\exp(-c\tau)u(\tau)$, with fit parameters $a$, $b$, $c$ and with $u(\tau)$ the Heaviside function] to the experimental CSs position distribution over the roundtrip (Fig.\,\ref{fig4}e).
The period ($a$) of the optimal fit and decay ($c$) parameters matches very well the computed one (Fig.\,\ref{fig4}d), with only a small discrepancy for the shift ($b$), which we believe stems from the coupling between the Raman self-frequency shift, not accounted for in the theoretical analysis, and the CSs oscillations.
Nonetheless, this confirms that the separation between the leading soliton and the first stable point is shorter than a lattice unit of length $1/2\nu_b$.
This shift results in distortions of the equilibrium position from a perfect lattice behind each vacancy owing to the cumulative effect of many CSs. 
Consequently, in the experiments, CSs sit on a disordered one-dimensional lattice, with a paracrystalline-like structure in time governed by the number and position of vacancies (Fig.\,\ref{fig4}f)\,\cite{millane_diffraction_2000} (see also Supplemental Material), while CSs localization on a perfect crystal lattice would occur in vacancy-less patterns. 

\textit{Conclusion---}We investigated the interplay between Brillouin lasing and Kerr-comb formation in coherently driven, doubly resonant ring cavities. We introduced a unified mean-field model for the forward and backward waves that explicitly accounts for their nonlinear coupling through Brillouin scattering. This coupling is shown to induce an effective many-body interaction mechanism.
We experimentally demonstrated that it stabilizes individual CSs on a temporal lattice of period $\approx1/2\nu_b$, resulting in long-lived, robust arrangements that would not otherwise exist. 
We stress that this effect differs from the well-known acoustically induced interaction responsible for pulse bunching in mode-locked lasers\,\cite{pilipetskii_acoustic_1995} and driven fiber Kerr-cavities\,\cite{jang_ultraweak_2013}, which we also  
observed in our experiments, though only below the first Brillouin threshold.
However, the shape of the interaction potential from a single soliton, together with the presence of vacancies, gives rise to a paracrystalline time structure. It is characterized by small stochastic cumulative distortions from perfect lattice ordering, which results in a broadened RF beatnote at $2\nu_b$. 

Vacancy‑free CS patterns would result in distortion‑free periodic soliton crystals with low frequency jitter, arising from the mutual coupling of thousands of CSs, since the fundamental jitter scales as $1/\sqrt N$ for $N$ coupled pulses (by comparison, CSs coupled through soliton‑radiated waves, such as Kelly sidebands, would involve only a few tens of pulses.)
This property makes such states attractive for applications in low‑noise optical frequency combs and low‑noise radio‑frequency generation.  
Unlike other schemes, the reduced linewidth of the Brillouin laser is not directly exploited here\,\cite{baiBrillouinKerrSolitonFrequency2021}, as the solitons are driven by the pump laser. However, the $S_2$ Stokes wave induces CS oscillations at a frequency linked to the detuning, opening a route toward further stabilization of the driving laser via the resulting heterogeneous RF beat signal. Finally, our mean-field model closely matches the experimental observations and provide a unified theoretical framework to further study single and multiple Kerr combs in fiber and integrated ring resonators\,\cite{baiBrillouinKerrSolitonFrequency2021, do_self-stabilized_2021, zhangSolitonMicrocombsMultiplexing2023}.

\section{Acknowledgment}
We acknowledge funding by the Fonds National de la Recherche Scientifique – FNRS through projects EQP n° 40021523 and EOS n°40007560 (PULSE), the ERC project HIGHRES and the Agence Nationale de la Recherche Scientifique – ANR through the project COMBY. C. Simon is a Research Fellow of the Fonds de la Recherche Scientifique – FNRS.

\bibliography{referencesV2, supplemental_ref}

@article{amrani_universal_2011,
	title = {Universal soliton pattern formations in passively mode-locked fiber lasers},
	volume = {36},
	issn = {0146-9592, 1539-4794},
	url = {https://www.osapublishing.org/abstract.cfm?URI=ol-36-9-1545},
	doi = {10.1364/OL.36.001545},
	number = {9},
	urldate = {2021-10-15},
	journal = {Optics Letters},
	author = {Amrani, Foued and Salhi, Mohamed and Grelu, Philippe and Leblond, Hervé and Sanchez, François},
	month = may,
	year = {2011},
	pages = {1545},
}

@article{baiBrillouinKerrSolitonFrequency2021,
  title = {Brillouin-{{Kerr Soliton Frequency Combs}} in an {{Optical Microresonator}}},
  author = {Bai, Yan and Zhang, Menghua and Shi, Qi and Ding, Shulin and Qin, Yingchun and Xie, Zhenda and Jiang, Xiaoshun and Xiao, Min},
  year = {2021},
  month = feb,
  journal = {Physical Review Letters},
  volume = {126},
  number = {6},
  pages = {063901},
  publisher = {American Physical Society},
  doi = {10.1103/PhysRevLett.126.063901},
  urldate = {2024-10-10}
}

@article{brajeBrillouinEnhancedHyperparametricGeneration2009,
  title = {Brillouin-{{Enhanced Hyperparametric Generation}} of an {{Optical Frequency Comb}} in a {{Monolithic Highly Nonlinear Fiber Cavity Pumped}} by a Cw {{Laser}}},
  author = {Braje, Danielle and Hollberg, Leo and Diddams, Scott},
  year = {2009},
  month = may,
  journal = {Physical Review Letters},
  volume = {102},
  number = {19},
  pages = {193902},
  publisher = {American Physical Society},
  doi = {10.1103/PhysRevLett.102.193902},
  urldate = {2024-10-02}
}

@article{bunelBrillouininducedKerrFrequency2025,
  title = {Brillouin-Induced {{Kerr}} Frequency Comb in Normal Dispersion Fiber {{Fabry Perot}} Resonators},
  author = {Bunel, Thomas and Lumeau, Julien and Moreau, Antonin and Fernandez, Arnaud and Llopis, Olivier and Bourcier, Germain and Perego, Auro M. and Conforti, Matteo and Mussot, Arnaud},
  year = {2025},
  month = jun,
  journal = {Nature Communications},
  volume = {16},
  number = {1},
  pages = {5160},
  publisher = {Nature Publishing Group},
  issn = {2041-1723},
  doi = {10.1038/s41467-025-60261-y},
  urldate = {2025-07-18},
  copyright = {2025 The Author(s)},
  langid = {english}
}

@article{buttner_phase-locking_2014,
	title = {Phase-locking and {Pulse} {Generation} in {Multi}-{Frequency} {Brillouin} {Oscillator} via {Four} {Wave} {Mixing}},
	volume = {4},
	copyright = {2014 The Author(s)},
	issn = {2045-2322},
	url = {https://www.nature.com/articles/srep05032},
	doi = {10.1038/srep05032},
	number = {1},
	urldate = {2023-04-06},
	journal = {Scientific Reports},
	publisher = {Nature Publishing Group},
	author = {Büttner, Thomas F. S. and Kabakova, Irina V. and Hudson, Darren D. and Pant, Ravi and Poulton, Christopher G. and Judge, Alexander C. and Eggleton, Benjamin J.},
	month = may,
	year = {2014},
	note = {Number: 1},
	pages = {5032},
}

@article{coenUniversalScalingLaws2013,
  title = {Universal Scaling Laws of {{Kerr}} Frequency Combs},
  author = {Coen, St{\'e}phane and Erkintalo, Miro},
  year = {2013},
  month = jun,
  journal = {Optics Letters},
  volume = {38},
  number = {11},
  pages = {1790--1792},
  issn = {1539-4794},
  doi = {10.1364/OL.38.001790},
  urldate = {2021-03-16},
  copyright = {\&\#169; 2013 Optical Society of America},
  langid = {english}
}

@article{cole_soliton_2017,
	title = {Soliton crystals in {Kerr} resonators},
	volume = {11},
	issn = {1749-4885, 1749-4893},
	url = {https://www.nature.com/articles/s41566-017-0009-z},
	doi = {10.1038/s41566-017-0009-z},
	number = {10},
	urldate = {2024-04-02},
	journal = {Nature Photonics},
	author = {Cole, Daniel C. and Lamb, Erin S. and Del’Haye, Pascal and Diddams, Scott A. and Papp, Scott B.},
	month = oct,
	year = {2017},
	pages = {671--676},
}

@article{ConfortiFast2025,
author = {Conforti, Matteo and Bunel, Thomas and Perego, Auro M and Mussot, Arnaud},
doi = {10.1103/1cfh-8c16},
issn = {2469-9926},
journal = {Physical Review A},
month = {sep},
number = {3},
pages = {033526},
publisher = {American Physical Society},
title = {{Fast and accurate modeling of Kerr-Brillouin combs in Fabry-Perot resonators}},
url = {https://link.aps.org/doi/10.1103/1cfh-8c16},
volume = {112},
year = {2025}
}

@article{debutLinewidthNarrowingBrillouin2000,
  title = {Linewidth Narrowing in {{Brillouin}} Lasers: {{Theoretical}} Analysis},
  shorttitle = {Linewidth Narrowing in {{Brillouin}} Lasers},
  author = {Debut, Alexis and Randoux, St{\'e}phane and Zemmouri, Jaouad},
  year = {2000},
  month = jul,
  journal = {Physical Review A},
  volume = {62},
  number = {2},
  pages = {023803},
  publisher = {American Physical Society},
  doi = {10.1103/PhysRevA.62.023803},
  urldate = {2025-07-31}
}

@article{deroh_dispersion_2023,
	title = {Dispersion engineering in a {Brillouin} fiber laser cavity for {Kerr} frequency comb formation},
	volume = {48},
	issn = {0146-9592, 1539-4794},
	url = {https://opg.optica.org/abstract.cfm?URI=ol-48-24-6388},
	doi = {10.1364/OL.506610},
	number = {24},
	urldate = {2024-06-27},
	journal = {Optics Letters},
	author = {Deroh, Moise and Lucas, Erwan and Kibler, Bertrand},
	month = dec,
	year = {2023},
	pages = {6388},
}

@article{do_self-stabilized_2021,
	title = {Self-stabilized soliton generation in a microresonator through mode-pulled {Brillouin} lasing},
	volume = {46},
	issn = {0146-9592, 1539-4794},
	url = {https://opg.optica.org/abstract.cfm?URI=ol-46-7-1772},
	doi = {10.1364/OL.419137},
	number = {7},
	urldate = {2023-03-21},
	journal = {Optics Letters},
	author = {Do, In Hwan and Kim, Dohyeong and Jeong, Dongin and Suk, Daewon and Kwon, Dohyeon and Kim, Jungwon and Lee, Jae Hoon and Lee, Hansuek},
	month = apr,
	year = {2021},
	pages = {1772},
}

@article{dongUnifiedApproachCascaded2016,
  title = {Unified Approach to Cascaded Stimulated {{Brillouin}} Scattering and Frequency-Comb Generation},
  author = {Dong, Mark and Winful, Herbert G.},
  year = {2016},
  month = apr,
  journal = {Physical Review A},
  volume = {93},
  number = {4},
  pages = {043851},
  publisher = {American Physical Society},
  doi = {10.1103/PhysRevA.93.043851},
  urldate = {2024-10-10}
}

@article{eggleton_brillouin_2019,
	title = {Brillouin integrated photonics},
	volume = {13},
	issn = {1749-4885, 1749-4893},
	url = {https://www.nature.com/articles/s41566-019-0498-z},
	doi = {10.1038/s41566-019-0498-z},
	number = {10},
	urldate = {2026-02-02},
	journal = {Nature Photonics},
	author = {Eggleton, Benjamin J. and Poulton, Christopher G. and Rakich, Peter T. and Steel, Michael. J. and Bahl, Gaurav},
	month = oct,
	year = {2019},
	pages = {664--677},
}

@article{englebert_parametrically_2021,
	title = {Parametrically driven {Kerr} cavity solitons},
	volume = {15},
	issn = {1749-4885, 1749-4893},
	url = {https://www.nature.com/articles/s41566-021-00858-z},
	doi = {10.1038/s41566-021-00858-z},
	number = {11},
	urldate = {2022-01-28},
	journal = {Nature Photonics},
	author = {Englebert, Nicolas and De Lucia, Francesco and Parra-Rivas, Pedro and Arabí, Carlos Mas and Sazio, Pier-John and Gorza, Simon-Pierre and Leo, François},
	month = nov,
	year = {2021},
	pages = {857--861},
}

@article{erkintalo_bunching_2015,
	title = {Bunching of temporal cavity solitons via forward {Brillouin} scattering},
	volume = {17},
	issn = {1367-2630},
	url = {https://iopscience.iop.org/article/10.1088/1367-2630/17/11/115009},
	doi = {10.1088/1367-2630/17/11/115009},
	number = {11},
	urldate = {2021-05-21},
	journal = {New Journal of Physics},
	author = {Erkintalo, Miro and Luo, Kathy and Jang, Jae K and Coen, Stéphane and Murdoch, Stuart G},
	month = nov,
	year = {2015},
	pages = {115009},
}

@article{erkintaloPhaseIntensityControl2022,
  title = {Phase and Intensity Control of Dissipative {{Kerr}} Cavity Solitons},
  author = {Erkintalo, Miro and Murdoch, Stuart G. and Coen, St{\'e}phane},
  year = {2022},
  month = mar,
  journal = {Journal of the Royal Society of New Zealand},
  volume = {52},
  number = {2},
  pages = {149--167},
  publisher = {Taylor \& Francis},
  issn = {0303-6758},
  doi = {10.1080/03036758.2021.1900296},
  urldate = {2025-07-24}
}

@article{fortier20YearsDevelopments2019,
  title = {20 Years of Developments in Optical Frequency Comb Technology and Applications},
  author = {Fortier, Tara and Baumann, Esther},
  year = {2019},
  month = dec,
  journal = {Communications Physics},
  volume = {2},
  number = {1},
  pages = {153},
  publisher = {Nature Publishing Group},
  issn = {2399-3650},
  doi = {10.1038/s42005-019-0249-y},
  urldate = {2025-09-02},
  copyright = {2019 This is a U.S. government work and not under copyright protection in the U.S.; foreign copyright protection may apply},
  langid = {english}
}

@article{gengHighlyStableLownoise2006,
  title = {Highly Stable Low-Noise {{Brillouin}} Fiber Laser with Ultranarrow Spectral Linewidth},
  author = {Geng, Jihong and Staines, S. and Wang, Zuolan and Zong, Jie and Blake, M. and Jiang, Shibin},
  year = {2006},
  month = sep,
  journal = {IEEE Photonics Technology Letters},
  volume = {18},
  number = {17},
  pages = {1813--1815},
  issn = {1941-0174},
  doi = {10.1109/LPT.2006.881145},
  urldate = {2024-10-11}
}

@article{grelu_dissipative_2012,
	title = {Dissipative solitons for mode-locked lasers},
	volume = {6},
	issn = {1749-4885, 1749-4893},
	url = {http://www.nature.com/articles/nphoton.2011.345},
	doi = {10.1038/nphoton.2011.345},
	number = {2},
	urldate = {2021-06-21},
	journal = {Nature Photonics},
	author = {Grelu, Philippe and Akhmediev, Nail},
	month = feb,
	year = {2012},
	pages = {84--92},
}

@article{gundavarapuSubhertzFundamentalLinewidth2019,
  title = {Sub-Hertz Fundamental Linewidth Photonic Integrated {{Brillouin}} Laser},
  author = {Gundavarapu, Sarat and Brodnik, Grant M. and Puckett, Matthew and Huffman, Taran and Bose, Debapam and Behunin, Ryan and Wu, Jianfeng and Qiu, Tiequn and Pinho, C{\'a}tia and Chauhan, Nitesh and Nohava, Jim and Rakich, Peter T. and Nelson, Karl D. and Salit, Mary and Blumenthal, Daniel J.},
  year = {2019},
  month = jan,
  journal = {Nature Photonics},
  volume = {13},
  number = {1},
  pages = {60--67},
  publisher = {Nature Publishing Group},
  issn = {1749-4893},
  doi = {10.1038/s41566-018-0313-2},
  urldate = {2024-10-10},
  copyright = {2018 The Author(s), under exclusive licence to Springer Nature Limited},
  langid = {english}
}

@article{heffernan_brillouin_2024,
	title = {Brillouin laser-driven terahertz oscillator up to 3 {THz} with femtosecond-level timing jitter},
	volume = {18},
	issn = {1749-4885, 1749-4893},
	url = {https://www.nature.com/articles/s41566-024-01513-z},
	doi = {10.1038/s41566-024-01513-z},
	number = {12},
	urldate = {2025-05-13},
	journal = {Nature Photonics},
	author = {Heffernan, Brendan M. and Greenberg, James and Hori, Takashi and Tanigawa, Tatsuya and Rolland, Antoine},
	month = dec,
	year = {2024},
	pages = {1263--1268},
}

@article{herrTemporalSolitonsOptical2014,
  title = {Temporal Solitons in Optical Microresonators},
  author = {Herr, Tobias and Brasch, Victor and Jost, John D. and Wang, Christine Y. and Kondratiev, Nikita M. and Gorodetsky, Michael L. and Kippenberg, Tobias J.},
  year = {2014},
  journal = {Nature Photonics},
  volume = {8},
  number = {2},
  pages = {145--152},
  publisher = {Nature Publishing Group UK London},
  urldate = {2024-10-14}
}

@article{huangTemporalSolitonOptical2019,
  title = {Temporal Soliton and Optical Frequency Comb Generation in a {{Brillouin}} Laser Cavity},
  author = {Huang, Yali and Li, Qing and Han, Junyuan and Jia, Zhixu and Yu, Yongsen and Yang, Yuede and Xiao, Jinlong and Wu, Jiliang and Zhang, Daming and Huang, Yongzhen and Qin, Weiping and Qin, Guanshi},
  year = {2019},
  month = dec,
  journal = {Optica},
  volume = {6},
  number = {12},
  pages = {1491--1497},
  publisher = {Optica Publishing Group},
  issn = {2334-2536},
  doi = {10.1364/OPTICA.6.001491},
  urldate = {2024-10-10},
  copyright = {{\copyright} 2019 Optical Society of America},
  langid = {english}
}

@article{jang_ultraweak_2013,
	title = {Ultraweak long-range interactions of solitons observed over astronomical distances},
	volume = {7},
	issn = {1749-4885, 1749-4893},
	url = {http://www.nature.com/articles/nphoton.2013.157},
	doi = {10.1038/nphoton.2013.157},
	number = {8},
	urldate = {2021-02-26},
	journal = {Nature Photonics},
	author = {Jang, Jae K. and Erkintalo, Miro and Murdoch, Stuart G. and Coen, Stéphane},
	month = aug,
	year = {2013},
	pages = {657--663},
}

@article{jia_photonic_2020,
	title = {Photonic {Flywheel} in a {Monolithic} {Fiber} {Resonator}},
	volume = {125},
	issn = {0031-9007, 1079-7114},
	url = {https://link.aps.org/doi/10.1103/PhysRevLett.125.143902},
	doi = {10.1103/PhysRevLett.125.143902},
	number = {14},
	urldate = {2024-05-02},
	journal = {Physical Review Letters},
	author = {Jia, Kunpeng and Wang, Xiaohan and Kwon, Dohyeon and Wang, Jiarong and Tsao, Eugene and Liu, Huaying and Ni, Xin and Guo, Jian and Yang, Mufan and Jiang, Xiaoshun and Kim, Jungwon and Zhu, Shi-ning and Xie, Zhenda and Huang, Shu-Wei},
	month = oct,
	year = {2020},
	pages = {143902},
}

@article{kobyakovStimulatedBrillouinScattering2010,
  title = {Stimulated {{Brillouin}} Scattering in Optical Fibers},
  author = {Kobyakov, Andrey and Sauer, Michael and Chowdhury, Dipak},
  year = {2010},
  month = mar,
  journal = {Advances in Optics and Photonics},
  volume = {2},
  number = {1},
  pages = {1--59},
  publisher = {Optica Publishing Group},
  issn = {1943-8206},
  doi = {10.1364/AOP.2.000001},
  urldate = {2024-10-11},
  copyright = {{\copyright} 2009 Optical Society of America},
  langid = {english}
}

@article{korobko_detuning_2020,
	title = {Detuning effects in {Brillouin} ring microresonator laser},
	volume = {28},
	issn = {1094-4087},
	url = {https://opg.optica.org/abstract.cfm?URI=oe-28-4-4962},
	doi = {10.1364/OE.382357},
	number = {4},
	urldate = {2022-11-03},
	journal = {Optics Express},
	author = {Korobko, D. A. and Zolotovskii, I. O. and Svetukhin, V. V. and Zhukov, A. V. and Fomin, A. N. and Borisova, C. V. and Fotiadi, A. A.},
	month = feb,
	year = {2020},
	pages = {4962},
}

@article{leoTemporalCavitySolitons2010,
  title = {Temporal Cavity Solitons in One-Dimensional {{Kerr}} Media as Bits in an All-Optical Buffer},
  author = {Leo, Fran{\c c}ois and Coen, St{\'e}phane and Kockaert, Pascal and Gorza, Simon-Pierre and Emplit, Philippe and Haelterman, Marc},
  year = {2010},
  month = jul,
  journal = {Nature Photonics},
  volume = {4},
  number = {7},
  pages = {471--476},
  publisher = {Nature Publishing Group},
  issn = {1749-4893},
  doi = {10.1038/nphoton.2010.120},
  urldate = {2024-10-14},
  copyright = {2010 Springer Nature Limited},
  langid = {english}
}

@article{li_optical_2017,
	title = {Optical frequency combs generated by four-wave mixing in a dual wavelength {Brillouin} laser cavity},
	volume = {7},
	url = {https://aip.scitation.org/doi/10.1063/1.4994861},
	doi = {10.1063/1.4994861},
	number = {7},
	urldate = {2023-03-21},
	journal = {AIP Advances},
	publisher = {American Institute of Physics},
	author = {Li, Qing and Jia, Zhi-xu and Li, Zhen-rui and Yang, Yue-de and Xiao, Jin-long and Chen, Shao-wu and Qin, Guan-shi and Huang, Yong-zhen and Qin, Wei-ping},
	month = jul,
	year = {2017},
	pages = {075215},
}

@article{liCharacterizationHighCoherence2012,
  title = {Characterization of a High Coherence, {{Brillouin}} Microcavity Laser on Silicon},
  author = {Li, Jiang and Lee, Hansuek and Chen, Tong and Vahala, Kerry J.},
  year = {2012},
  month = aug,
  journal = {Optics Express},
  volume = {20},
  number = {18},
  pages = {20170--20180},
  publisher = {Optica Publishing Group},
  issn = {1094-4087},
  doi = {10.1364/OE.20.020170},
  urldate = {2024-10-11},
  copyright = {{\copyright} 2012 OSA},
  langid = {english}
}

@article{liMicroresonatorBrillouinGyroscope2017,
  title = {Microresonator {{Brillouin}} Gyroscope},
  author = {Li, Jiang and Suh, Myoung-Gyun and Vahala, Kerry},
  year = {2017},
  month = mar,
  journal = {Optica},
  volume = {4},
  number = {3},
  pages = {346--348},
  publisher = {Optica Publishing Group},
  issn = {2334-2536},
  doi = {10.1364/OPTICA.4.000346},
  urldate = {2024-10-11},
  copyright = {{\copyright} 2017 Optical Society of America},
  langid = {english}
}

@article{liMicrowaveSynthesizerUsing2013,
  title = {Microwave Synthesizer Using an On-Chip {{Brillouin}} Oscillator},
  author = {Li, Jiang and Lee, Hansuek and Vahala, Kerry J.},
  year = {2013},
  month = jun,
  journal = {Nature Communications},
  volume = {4},
  number = {1},
  pages = {2097},
  publisher = {Nature Publishing Group},
  issn = {2041-1723},
  doi = {10.1038/ncomms3097},
  urldate = {2024-10-10},
  copyright = {2013 The Author(s)},
  langid = {english}
}

@article{liu_photonic_2020,
	title = {Photonic microwave generation in the {X}- and {K}-band using integrated soliton microcombs},
	volume = {14},
	issn = {1749-4885, 1749-4893},
	url = {https://www.nature.com/articles/s41566-020-0617-x},
	doi = {10.1038/s41566-020-0617-x},
	number = {8},
	urldate = {2025-03-18},
	journal = {Nature Photonics},
	author = {Liu, Junqiu and Lucas, Erwan and Raja, Arslan S. and He, Jijun and Riemensberger, Johann and Wang, Rui Ning and Karpov, Maxim and Guo, Hairun and Bouchand, Romain and Kippenberg, Tobias J.},
	month = aug,
	year = {2020},
	pages = {486--491},
}

@article{loh_operation_2020,
	title = {Operation of an optical atomic clock with a {Brillouin} laser subsystem},
	volume = {588},
	issn = {0028-0836, 1476-4687},
	url = {http://www.nature.com/articles/s41586-020-2981-6},
	doi = {10.1038/s41586-020-2981-6},
	number = {7837},
	urldate = {2022-12-14},
	journal = {Nature},
	author = {Loh, William and Stuart, Jules and Reens, David and Bruzewicz, Colin D. and Braje, Danielle and Chiaverini, John and Juodawlkis, Paul W. and Sage, Jeremy M. and McConnell, Robert},
	month = dec,
	year = {2020},
	pages = {244--249},
}

@article{loh_ultra-narrow_2019,
	title = {Ultra-narrow linewidth {Brillouin} laser with nanokelvin temperature self-referencing},
	volume = {6},
	copyright = {© 2019 Optical Society of America},
	issn = {2334-2536},
	url = {https://opg.optica.org/optica/abstract.cfm?uri=optica-6-2-152},
	doi = {10.1364/OPTICA.6.000152},
	number = {2},
	urldate = {2026-01-16},
	journal = {Optica},
	publisher = {Optica Publishing Group},
	author = {Loh, William and Yegnanarayanan, Siva and O’Donnell, Frederick and Juodawlkis, Paul W.},
	month = feb,
	year = {2019},
	pages = {152--159},
}

@article{lucasDynamicInterplayKerr2023,
  title = {Dynamic {{Interplay Between Kerr Combs}} and {{Brillouin Lasing}} in {{Fiber Cavities}}},
  author = {Lucas, Erwan and Deroh, Moise and Kibler, Bertrand},
  year = {2023},
  journal = {Laser \& Photonics Reviews},
  volume = {17},
  number = {12},
  pages = {2300041},
  issn = {1863-8899},
  doi = {10.1002/lpor.202300041},
  urldate = {2024-10-11},
  copyright = {{\copyright} 2023 Wiley-VCH GmbH},
  langid = {english}
}

@article{lugiato_spatial_1987,
	title = {Spatial {Dissipative} {Structures} in {Passive} {Optical} {Systems}},
	volume = {58},
	url = {https://link.aps.org/doi/10.1103/PhysRevLett.58.2209},
	doi = {10.1103/PhysRevLett.58.2209},
	number = {21},
	urldate = {2022-05-10},
	journal = {Physical Review Letters},
	publisher = {American Physical Society},
	author = {Lugiato, L. A. and Lefever, R.},
	month = may,
	year = {1987},
	pages = {2209--2211},
}

@article{maggipintoCavitySolitonsSemiconductor2000,
  title = {Cavity Solitons in Semiconductor Microresonators:   {{Existence}}, Stability, and Dynamical Properties},
  shorttitle = {Cavity Solitons in Semiconductor Microresonators},
  author = {Maggipinto, T. and Brambilla, M. and Harkness, G. K. and Firth, W. J.},
  year = {2000},
  month = dec,
  journal = {Physical Review E},
  volume = {62},
  number = {6},
  pages = {8726--8739},
  publisher = {American Physical Society},
  doi = {10.1103/PhysRevE.62.8726},
  urldate = {2025-09-01}
}

@article{marin-palomo_microresonator-based_2017,
	title = {Microresonator-based solitons for massively parallel coherent optical communications},
	volume = {546},
	issn = {0028-0836, 1476-4687},
	url = {http://www.nature.com/articles/nature22387},
	doi = {10.1038/nature22387},
	number = {7657},
	urldate = {2020-05-07},
	journal = {Nature},
	author = {Marin-Palomo, Pablo and Kemal, Juned N. and Karpov, Maxim and Kordts, Arne and Pfeifle, Joerg and Pfeiffer, Martin H. P. and Trocha, Philipp and Wolf, Stefan and Brasch, Victor and Anderson, Miles H. and Rosenberger, Ralf and Vijayan, Kovendhan and Freude, Wolfgang and Kippenberg, Tobias J. and Koos, Christian},
	month = jun,
	year = {2017},
	pages = {274--279},
}

@article{millane_diffraction_2000,
	title = {Diffraction by one-dimensional paracrystals and perturbed lattices},
	volume = {56},
	issn = {0108-7673},
	url = {https://journals.iucr.org/paper?S0108767300008138},
	doi = {10.1107/S0108767300008138},
	number = {5},
	urldate = {2026-01-07},
	journal = {Acta Crystallographica Section A Foundations of Crystallography},
	author = {Millane, R. P. and Eads, J. L.},
	month = sep,
	year = {2000},
	pages = {497--506},
}

@article{murrayDistributedBrillouinFiber2022,
  title = {Distributed {{Brillouin}} Fiber Laser Sensor},
  author = {Murray, Joseph B. and Cerjan, Alex and Redding, Brandon},
  year = {2022},
  month = jan,
  journal = {Optica},
  volume = {9},
  number = {1},
  pages = {80--87},
  publisher = {Optica Publishing Group},
  issn = {2334-2536},
  doi = {10.1364/OPTICA.435716},
  urldate = {2024-10-11},
  copyright = {{\copyright} 2022 Optical Society of America},
  langid = {english}
}

@article{newman_architecture_2019,
	title = {Architecture for the photonic integration of an optical atomic clock},
	volume = {6},
	copyright = {\&\#169; 2019 Optical Society of America},
	issn = {2334-2536},
	url = {https://opg.optica.org/optica/abstract.cfm?uri=optica-6-5-680},
	doi = {10.1364/OPTICA.6.000680},
	number = {5},
	urldate = {2022-04-29},
	journal = {Optica},
	publisher = {Optica Publishing Group},
	author = {Newman, Zachary L. and Maurice, Vincent and Drake, Tara and Stone, Jordan R. and Briles, Travis C. and Spencer, Daryl T. and Fredrick, Connor and Li, Qing and Westly, Daron and Ilic, B. R. and Shen, Boqiang and Suh, Myoung-Gyun and Yang, Ki Youl and Johnson, Cort and Johnson, David M. S. and Hollberg, Leo and Vahala, Kerry J. and Srinivasan, Kartik and Diddams, Scott A. and Kitching, John and Papp, Scott B. and Hummon, Matthew T.},
	month = may,
	year = {2019},
	pages = {680--685},
}

@article{nimmesgern_soliton_2021,
	title = {Soliton molecules in femtosecond fiber lasers: universal binding mechanism and direct electronic control},
	volume = {8},
	issn = {2334-2536},
	shorttitle = {Soliton molecules in femtosecond fiber lasers},
	url = {https://opg.optica.org/abstract.cfm?URI=optica-8-10-1334},
	doi = {10.1364/OPTICA.439905},
	number = {10},
	urldate = {2025-05-06},
	journal = {Optica},
	author = {Nimmesgern, Luca and Beckh, Cornelius and Kempf, Hannes and Leitenstorfer, Alfred and Herink, Georg},
	month = oct,
	year = {2021},
	pages = {1334},
}

@article{pilipetskii_acoustic_1995,
	title = {Acoustic effect in passively mode-locked fiber ring lasers},
	volume = {20},
	copyright = {https://doi.org/10.1364/OA\_License\_v1\#VOR},
	issn = {0146-9592, 1539-4794},
	url = {https://opg.optica.org/abstract.cfm?URI=ol-20-8-907},
	doi = {10.1364/OL.20.000907},
	number = {8},
	urldate = {2025-10-21},
	journal = {Optics Letters},
	author = {Pilipetskii, A. N. and Golovchenko, E. A. and Menyuk, C. R.},
	month = apr,
	year = {1995},
	pages = {907},
}

@article{riemensberger_massively_2020,
	title = {Massively parallel coherent laser ranging using a soliton microcomb},
	volume = {581},
	copyright = {2020 The Author(s), under exclusive licence to Springer Nature Limited},
	issn = {1476-4687},
	url = {https://www.nature.com/articles/s41586-020-2239-3},
	doi = {10.1038/s41586-020-2239-3},
	number = {7807},
	urldate = {2022-11-15},
	journal = {Nature},
	publisher = {Nature Publishing Group},
	author = {Riemensberger, Johann and Lukashchuk, Anton and Karpov, Maxim and Weng, Wenle and Lucas, Erwan and Liu, Junqiu and Kippenberg, Tobias J.},
	month = may,
	year = {2020},
	note = {Number: 7807},
	pages = {164--170},
}

@article{tang_stable_2011,
	title = {A stable optical comb with double-{Brillouin}-frequency spacing assisted by multiple four-wave mixing processes},
	volume = {17},
	issn = {1068-5200},
	url = {https://www.sciencedirect.com/science/article/pii/S1068520011001118},
	doi = {10.1016/j.yofte.2011.08.005},
	number = {6},
	urldate = {2023-04-06},
	journal = {Optical Fiber Technology},
	author = {Tang, Jianguan and Sun, Junqiang and Chen, Tao and Zhou, Yujie},
	month = dec,
	year = {2011},
	pages = {608--611},
}

@article{wang_universal_2017,
	title = {Universal mechanism for the binding of temporal cavity solitons},
	volume = {4},
	issn = {2334-2536},
	url = {https://www.osapublishing.org/abstract.cfm?URI=optica-4-8-855},
	doi = {10.1364/OPTICA.4.000855},
	number = {8},
	urldate = {2020-05-07},
	journal = {Optica},
	author = {Wang, Yadong and Leo, François and Fatome, Julien and Erkintalo, Miro and Murdoch, Stuart G. and Coen, Stéphane},
	month = aug,
	year = {2017},
	pages = {855},
}

@article{weng_heteronuclear_2020,
	title = {Heteronuclear soliton molecules in optical microresonators},
	volume = {11},
	issn = {2041-1723},
	url = {https://www.nature.com/articles/s41467-020-15720-z},
	doi = {10.1038/s41467-020-15720-z},
	number = {1},
	urldate = {2026-01-29},
	journal = {Nature Communications},
	author = {Weng, Wenle and Bouchand, Romain and Lucas, Erwan and Obrzud, Ewelina and Herr, Tobias and Kippenberg, Tobias J.},
	month = may,
	year = {2020},
	pages = {2402},
}

@article{zarinetchiStimulatedBrillouinFiberoptic1991,
  title = {Stimulated {{Brillouin}} Fiber-Optic Laser Gyroscope},
  author = {Zarinetchi, F. and Smith, S. P. and Ezekiel, S.},
  year = {1991},
  month = feb,
  journal = {Optics Letters},
  volume = {16},
  number = {4},
  pages = {229--231},
  publisher = {Optica Publishing Group},
  issn = {1539-4794},
  doi = {10.1364/OL.16.000229},
  urldate = {2024-10-11},
  copyright = {{\copyright} 1991 Optical Society of America},
  langid = {english}
}

@article{zhangSolitonMicrocombsMultiplexing2023,
  title = {Soliton {{Microcombs Multiplexing Using Intracavity-Stimulated Brillouin Lasers}}},
  author = {Zhang, Hao and Tan, Teng and Chen, Hao-Jing and Yu, Yan and Wang, Wenting and Chang, Bing and Liang, Yupei and Guo, Yanhong and Zhou, Heng and Xia, Handing and Gong, Qihuang and Wong, Chee Wei and Rao, Yunjiang and Xiao, Yun-Feng and Yao, Baicheng},
  year = {2023},
  month = apr,
  journal = {Physical Review Letters},
  volume = {130},
  number = {15},
  pages = {153802},
  publisher = {American Physical Society},
  doi = {10.1103/PhysRevLett.130.153802},
  urldate = {2024-10-11}
}

@book{lugiato_prati_brambilla_2015,
  author    = {Lugiato, L. A. and Prati, F. and Brambilla, M.},
  title     = {Nonlinear Optical Systems},
  publisher = {Cambridge University Press},
  year      = {2015},
  doi       = {10.1017/CBO9781107477254}
}

@article{Ziani2024,
  author  = {Ziani, Z. and Bunel, T. and Perego, A. M. and Mussot, A. and Conforti, M.},
  title   = {Theory of modulation instability in Kerr Fabry-Perot resonators beyond the mean-field limit},
  journal = {Physical Review A},
  volume  = {109},
  pages   = {013507},
  year    = {2024},
  doi     = {10.1103/PhysRevA.109.013507}
}

@article{Chembo2013,
  author  = {Chembo, Y. K. and Menyuk, C. R.},
  title   = {Spatiotemporal Lugiato-Lefever formalism for Kerr-comb generation in whispering-gallery-mode resonators},
  journal = {Physical Review A},
  volume  = {87},
  pages   = {053852},
  year    = {2013},
  doi     = {10.1103/PhysRevA.87.053852}
}

@article{jangAllopticalBufferBased2016,
  title = {All-Optical Buffer Based on Temporal Cavity Solitons Operating at 10\,\,{{Gb}}/s},
  author = {Jang, Jae K. and Erkintalo, Miro and Schr{\"o}der, Jochen and Eggleton, Benjamin J. and Murdoch, Stuart G. and Coen, St{\'e}phane},
  year = {2016},
  month = oct,
  journal = {Optics Letters, Vol. 41, Issue 19, pp. 4526-4529},
  publisher = {Optica Publishing Group},
  doi = {10.1364/OL.41.004526},
  urldate = {2025-07-25},
  copyright = {{\copyright} 2016 Optical Society of America},
  langid = {english}
}

@article{hultFourthOrderRungeKutta2007,
  title = {A {{Fourth-Order Runge}}--{{Kutta}} in the {{Interaction Picture Method}} for {{Simulating Supercontinuum Generation}} in {{Optical Fibers}}},
  author = {Hult, Johan},
  year = {2007},
  month = dec,
  journal = {Journal of Lightwave Technology},
  volume = {25},
  number = {12},
  pages = {3770--3775},
  issn = {1558-2213},
  doi = {10.1109/JLT.2007.909373},
  urldate = {2025-07-18}
}

@article{Firth1981,
  author  = {Firth, W.},
  title   = {Stability of nonlinear Fabry-Perot resonators},
  journal = {Optics Communications},
  volume  = {39},
  pages   = {343--},
  year    = {1981},
  doi     = {10.1016/0030-4018(81)90106-1}
}

@article{Firth2021,
  author  = {Firth, W. J. and Geddes, J. B. and Karst, N. J. and Oppo, G.-L.},
  title   = {Analytic instability thresholds in folded Kerr resonators of arbitrary finesse},
  journal = {Physical Review A},
  volume  = {103},
  pages   = {023510},
  year    = {2021},
  doi     = {10.1103/PhysRevA.103.023510}
}

\section*{End Matter}

\textit{Theoretical model}--Our theoretical analysis of the intracavity nonlinear dynamics, along with numerical simulations, relies on the following coupled mean-field equations:  

\begin{align}
  T_{r}&\frac{\partial\psi_{F/B}}{\partial t} = -\left(\frac{\Lambda}{2}+i\delta\right)\psi_{F/B}-i \frac{\beta_{2}L}{2\beta_{1}^{2}}\frac{\partial^{2}\psi_{F/B}}{\partial z^{2}}\notag \\
  &+\sqrt{\theta P_\mathrm{in}^{F/B}}+i\gamma L\left(|\psi_{F/B}|^{2}+2\langle|\psi_{B/F}|^{2}\rangle \right)\psi_{F/B} \notag \\
  & +i\frac{g_{b}}{2A_\mathrm{eff}}\psi_{F/B}\circledast \left[\tilde{h}_b(\mp z)\times\left(\psi_{B/F}^{*}\circledast  \psi_{B/F}(-z)\right)\right], \label{eq1}\\
  &h_b(z) =\frac{\beta_{1}\Omega_{b}\Gamma_{b}}{\sqrt{\Omega_{b}^{2}-(\Gamma_{b}/2)^{2}}}\exp \left(-\frac{\Gamma_{b}}{2}\beta_{1}z\right) \times \notag \\
  & \sin\left(\sqrt{\Omega_{b}^{2}-(\Gamma_{b}/2)^{2}}\,\beta_{1}z\right) u(z), \label{eq2}
\end{align}
where $t$ and $z$ are the temporal and spatial coordinates in a reference frame moving at the light velocity in the cavity, and $T_r$ is the round-trip time. $\psi_{F/B}$ are the envelopes of the forward $(F)$ and backward $(B)$ intracavity electric fields, with $|\psi_{F/B}|^2$ their powers. $u(z)$ is the Heaviside step function, $\langle a \rangle$ is the spatial average of the quantity $a$, $\circledast$ is the spatial periodic convolution operator, and
$\tilde{h}_b(z) = \sum_{n=-\infty}^\infty h_b(z - nL)$ is the periodic replication of the Brillouin response $h_b(z)$, with $L$ the cavity length. 
$P_\mathrm{in}^{F/B}$ are the input powers (note that only $ \psi_{F}$ is driven in the experiments, hence $P_\mathrm{in}^{B}=0$). 
$\delta$ is the detuning from the closest cavity resonance; $\beta_{1,2}$ are the inverse of the group velocity and group velocity dispersion, respectively; $\theta$ is the input coupling ratio, and $\gamma$ is the nonlinear Kerr coefficient. 
The Brillouin gain is characterized by its strength $g_b/A_\mathrm{eff}$, its bandwidth $\Gamma_b$, and the Stokes shift $\Omega_b$.   
This set of equations is derived from the model presented in\,\cite{ConfortiFast2025} for Fabry–Perot cavities, and adapted here to ring resonators where the forward and backward fields are distinct (see Supplemental Material\,). We note that since the FSR of our cavity ($\qty{6.685}{\mega \hertz}$) is several times smaller than the Brillouin bandwidth, $\Gamma_b/2\pi \approx \qty{28}{\mega \hertz}$\,\cite{kobyakovStimulatedBrillouinScattering2010}, $\tilde{h}_b(z)$ can be simplified to $h_b(z)$.

To theoretically analyze the interaction of a single (forward) CS with the backward wave $S_1$,
assuming single-mode backward Stokes lasing, we first note that the bandwidth of the forward gain induced by $\psi_B=S_1 \exp(i\Omega_b \beta_1 z)$ is much narrower than the soliton spectrum ($\sim \si{\tera \hertz}$). 
Consequently, the contribution of a single CS to $\psi_F$ in the convolution term in Eq.(\ref{eq1}) can be approximated by a Dirac-delta function centered on the CS position $z_\mathrm{CS}$ and scaled by the integral of the (sech-shaped) soliton field $\pi \sqrt{P_\mathrm{CS}} \tau_\mathrm{CS} / \beta_1 \exp( i \phi_\mathrm{CS})$, with $P_\mathrm{CS}$, $\tau_\mathrm{CS}$, and $\phi_\mathrm{CS}$ denoting the soliton peak power, duration, and phase, respectively.
The Brillouin interaction term between $S_1$ and the CS in the equation of the forward wave thus reads   
{$+i \Lambda \pi \sqrt{P_\mathrm{CS}} \tau_\mathrm{CS} /(2\beta_1^{-1}) \exp( i \phi_\mathrm{CS})\times {h}_b(z_\mathrm{CS} - z) \exp\left[i\Omega_b \beta_1 (z_\mathrm{CS} - z)\right]$},
in which we made the substitution corresponding to the lasing threshold of $S_2$, which requires $g_{b} L|S_1|^2/A_\mathrm{eff} = \Lambda$ [at high detuning, the thresholds of the backward $S_1$ and forward $S_2$ Stokes waves are crossed simultaneously for increasing $P_\mathrm{in}^F$ in doubly resonant Brillouin lasers with $\Gamma_b \gg 2\pi\times\mathrm{FSR}$ (see Supplemental Material\, Fig.S2(d)].\\

\textit{Experimental setup.}--The experiments are performed in a ring resonator which consists of a standard optical fiber (SMF-28) looped on itself by a 99/1 coupler. The resulting low-loss cavity is $\qty{30.7}{\meter}$-long and exhibits a round-trip loss of $\Lambda = 3.5\%$ (1$\%$ input coupling), corresponding to a finesse of $\mathcal{F} = 177$ ($2.55 \times 10^9$ Q-factor). It is coherently driven by a narrow-linewidth continuous-wave laser at 1550.8\,nm (see Supplemental Material\, for a detailed schematic of the experimental setup).

In the experiments, we either scan the laser frequency across the resonance or stabilize its detuning within the soliton region. 
For stabilization, we implement a dedicated scheme. We measure, on a logarithmic scale, the ratio between the optical powers in two adjacent frequency bands offset from the driving wavelength. This amounts to considering the slope of the soliton spectrum, uniquely determined by the linear cavity detuning\,\cite{coenUniversalScalingLaws2013}. 
The control thus does not require an additional laser signal and is independent of the number of stochastically generated cavity solitons.
The long-term stability of the soliton pattern is studied by implementing an FPGA-based scheme in which the resonance is first scanned and, once CSs are detected, the dual-log stabilization loop engages to lock the detuning.

When the intracavity power increases, it eventually crosses the Brillouin laser threshold $g_bLP_\mathrm{cav}/A_\mathrm{eff} = \Lambda$. 
We measure an input threshold power at resonance $P_\mathrm{in}^\mathrm{th} \approx 300$\,$\mu$W, which corresponds to $P_\mathrm{cav}^\mathrm{th} \approx \qty{11.5}{\milli\watt}$ and a Brillouin gain $g_b/A_\mathrm{eff} = \qty{0.1 \pm 0.03}{\per \watt \per\meter}$, 
in agreement with previous results in standard SMF fibers\, \cite{kobyakovStimulatedBrillouinScattering2010}.

\end{document}



\title{Supplementary information for: \\Self-organization of cavity solitons in Brillouin–Kerr ring resonators}

\author{Corentin Simon}
\affiliation{OPERA-photonics, Université libre de Bruxelles (U.L.B.), 50~Avenue F. D. Roosevelt, CP 194/5, B-1050 Brussels, Belgium}
\author{Matteo Conforti}
\affiliation{University of Lille, CNRS, UMR 8523-PhLAM Physique des Lasers, Atomes et Molécules, F-59000, Lille, France}
\author{Loïc Van Bellinghen}
\affiliation{OPERA-photonics, Université libre de Bruxelles (U.L.B.), 50~Avenue F. D. Roosevelt, CP 194/5, B-1050 Brussels, Belgium}
\author{Maxime Fournier}
\affiliation{OPERA-photonics, Université libre de Bruxelles (U.L.B.), 50~Avenue F. D. Roosevelt, CP 194/5, B-1050 Brussels, Belgium}
\author{Arnaud Mussot}
\affiliation{University of Lille, CNRS, UMR 8523-PhLAM Physique des Lasers, Atomes et Molécules, F-59000, Lille, France}
\author{François Leo}
\affiliation{OPERA-photonics, Université libre de Bruxelles (U.L.B.), 50~Avenue F. D. Roosevelt, CP 194/5, B-1050 Brussels, Belgium}
\author{Simon-Pierre Gorza}
\affiliation{OPERA-photonics, Université libre de Bruxelles (U.L.B.), 50~Avenue F. D. Roosevelt, CP 194/5, B-1050 Brussels, Belgium}

\date{5.~March 2026}

\begin{abstract}

  This supplementary information is organized as follows. In the first section we give a detailed description of the experimental setup. In the second section we derive the mean-field equations describing the evolution of the counter-propagating optical fields in the ring resonator under the action of Kerr and Brillouin nonlinearities.
  In the third section, we describe the mechanism that couples soliton vacancies to the lattice distortion.
  In the fourth section, we provide a comparison of the threshold for CSs formation and Brillouin lasing. 
  The set of parameters and the method used for the simulations are detailed in the fifth section. 
 An additional figure at the end presents further simulations of the resonator’s nonlinear dynamics.
\end{abstract}


\maketitle


   \section{I. Experimental setup}

    \begin{figure}[h]
      \includegraphics{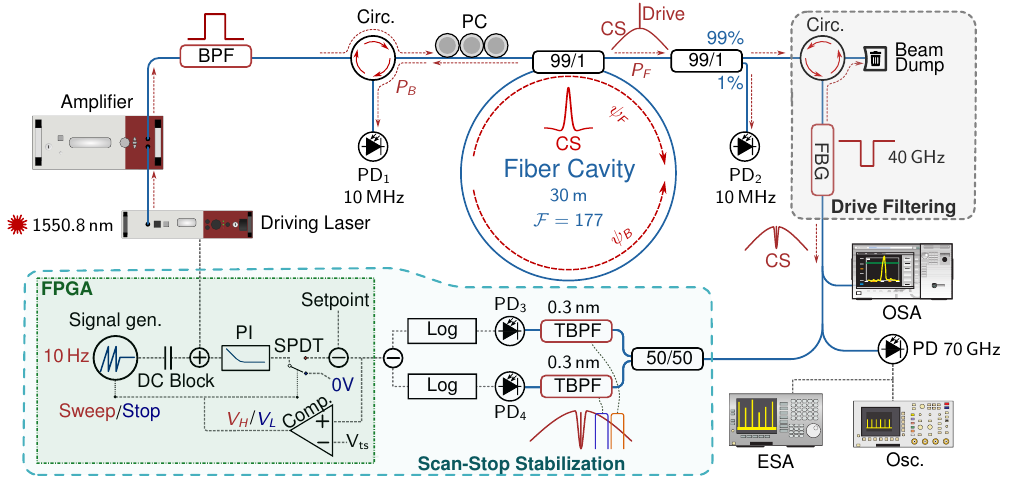}
      \caption{Experimental set-up. Cavity solitons (CSs) and Brillouin lasing are generated in a ring cavity consisting of a fiber loop closed on itself and a 99/1 optical coupler. 
      The driving beam is obtained from a continuous‑wave laser (Driving laser), followed by an amplifier and a band‑pass filter (BPF). 
      The signal in the through port is filtered by a fiber Bragg grating (FBG). This signal is then sent either to measurement apparatuses or to the FPGA-based feedback loop for stabilization purposes. Circ., circulator, PC, polarization controller, TBPF (tunable band-pass filter), PD, photodiode, SPDT, single pole double throw switch, PI, proportional-integral controller, OSA, optical spectrum analyzer, Osc., oscilloscope, ESA, electric spectrum analyzer. $\psi_F$ ($\psi_B$) forward (backward) intra-cavity field.
      }
      \label{fig-sup1:exp-setup}
    \end{figure}

Experiments were performed in a ring cavity made of $L= \qty{30.7}{ \meter}$ standard single-mode fiber [see Fig.\,\eqref{fig-sup1:exp-setup}]. The driving continuous-wave laser is a Koheras Adjustik at $\qty{1550.8}{\nano \meter}$. A commercial amplifier (Koheras Boostik) is used to amplify the driving signal power. The amplified spontaneous emission (ASE) generated in the EDFA is suppressed by a narrow bandpass filter ($\qty{100}{\giga\hertz}$), centered on the driving laser wavelength. The driving beam is launched into the cavity through the coupler ($\theta\approx1\%$) with a $\qty{600}{\milli\watt}$ input power. The polarization controller before the cavity aligns the input polarization with one of the two cavity eigenmodes, while the first circulator separates the driving beam from the counterpropagating Brillouin laser signal. 
Its power is monitored by the first photodiode PD$_1$. The signal at the cavity through port is first split into two. The 1\% tapped signal is used to measure the average power on a low-bandwidth photodiode (PD$_2$). The remaining part of the signal is then filtered by a narrowband fiber Bragg grating ($\qty{40}{\giga\hertz}$, $\qty{40}{\decibel}$ extinction ratio) to remove the driving frequency. Part of the filtered signal is sent to an optical spectrum analyzer and to a 70\,GHz photodiode for time measurement on an oscilloscope (80 GHz bandwidth, 256 gigasamples per second) or to a radio-frequency spectrum analyzer.\\

The stabilization of the detuning between the driving laser and the cavity was performed by implementing a proportional-integral (PI) feedback loop to lock the laser frequency. A dedicated dual-logarithmic detection scheme generates the error signal. This signal is based on the logarithm of the ratio of the optical powers measured by PD3 and PD4 in two adjacent filtered frequency bands.  
This is equivalent to measuring the slope of the soliton spectrum on a logarithmic scale, which is solely determined by the cavity detuning \cite{coenUniversalScalingLaws2013}. The resulting control signal thus enables detuning stabilization while remaining independent of the number of CSs, contrary to schemes based on a single frequency filter\,\cite{jangAllopticalBufferBased2016}.\\     

To determine the Brillouin gain (i.e. $g_b /A_\mathrm{eff}$) of our optical fiber, we measured the Brillouin lasing threshold input power $P_b^\mathrm{th}$ of our cavity.
To this end, we used a PDH locking scheme to drive the cavity on resonance, and we progressively increased the input power until a backward signal is detected. Taking into account the phase modulation of the driving laser for the PDH locking, the Brillouin gain is given by: 
    \begin{equation}
      (g_b /A_\mathrm{eff})^{-1} = \mathcal{F}^3 \theta P_b^\mathrm{th} L / (2\pi^3) \times J_0\left(\pi V_\mathrm{PDH} / V_\pi\right)^2,
    \end{equation}
where $\mathcal{F}=2\pi/\Lambda$ is the cavity finesse, $J_0$ denotes the Bessel function of the first kind, and $V_\mathrm{PDH}$, $V_\pi$ are the amplitude of the voltage signal sent to the modulator and the $V_\pi$ voltage of the modulator, respectively. We used $V_\mathrm{PDH} = \qty{0.5}{\volt}$ and $V_\pi = \qty{5\pm1}{\volt}$ (we note that this leads to a reduction of the driving power of 4.5\% compared to the unmodulated case).
We measured $P_b^\mathrm{th} = \qty{300 \pm 60}{\micro \watt}$. Thus, taking into account the values of the other parameters and their uncertainties: $\theta = \qty{0.013 \pm 0.003}{}$ and $\mathcal{F} = \qty{177 \pm 3}{}$, a Brillouin gain value of $g_b = \qty{0.1 \pm 0.03}{\per\watt\per\meter}$ is obtained.

    \newpage


\section{II. Derivation of the mean-field model}
The evolution equations for the slowly varying forward (clockwise) $F(z,t)$ and backward (counterclockwise) $B(z,t)$ optical fields, and the acoustic wave $Q$ read as\,\cite{ConfortiFast2025,dongUnifiedApproachCascaded2016,Firth1981,Firth2021}:
\begin{align}
  \frac{\partial F}{\partial z}+\beta_1\frac{\partial F}{\partial t} \label{propF}
  + i\frac{\beta_2}{2}\frac{\partial^2 F}{\partial t^2}&=i\gamma\left(|F|^2+X|B|^2\right)F+\frac{g_b}{2A_{\rm eff}}QB,\\
  -\frac{\partial B}{\partial z}+\beta_1\frac{\partial B}{\partial t} \label{propB}
  + i\frac{\beta_2}{2}\frac{\partial^2 B}{\partial t^2}&=i\gamma\left(|B|^2+X|F|^2\right)B-\frac{g_b}{2A_{\rm eff}}Q^*F,\\
  \left(\frac{\partial^2}{\partial t^2}+\Gamma_b\frac{\partial}{\partial t}+\Omega_b^2\right)Q&=i\Omega_b\Gamma_b FB^*,\label{Q}
\end{align}
where $v_g=\beta_1^{-1}$ is the group velocity, $\beta_2$  the group velocity  dispersion (GVD), and $\gamma$  the waveguide nonlinear coefficient (m$^{-1}$W$^{-1}$). The parameter $X$ describes the cross-phase modulation (XPM) and is taken to be $X=2$. The Brillouin gain is $g_b$ (m/W), $A_{\rm eff}$ is the effective area of the mode, $\tau_b=1/\Gamma_b$ is the phonon lifetime and $\Omega_b$ is the Brillouin shift.
The optical fields are normalized in such a way that their square modulus gives the instantaneous power.

Assuming a ring resonator and external coherent driving, the propagation equations (\ref{propF},\ref{propB}) are complemented with the following boundary conditions at the coupler, which connect the beginning ($z=0$) and the end ($z=L$) of the resonator:
\begin{align}
  F(0,t)&=\mathcal{T} E_{\rm in} + \mathcal{R} e^{-i\delta}F(L,t),\label{BC1}\\
  B(L,t)&=\mathcal{T} E_{\rm in}^B+ \mathcal{R} e^{-i\delta} B(0,t),\label{BC2}
\end{align}
where $\delta=2\pi m_0-\phi_0$ is the phase detuning between the driving fields  ($E_{\rm in}$, $E_{\rm in}^B$) and the closest resonance indexed by the integer $m_0$ (assumed here identical for both propagation directions for simplicity).
The phase $\phi_0$ accounts for the phase acquired during propagation over the resonator length $L$ ($\beta_0 L$, with $\beta_0$ the propagation constant of the mode), and any possible contribution from the coupler.
The coupler transmission and reflection coefficients in amplitude are $\mathcal{T}$ and $\mathcal{R}$, respectively.\\

To develop a mean-field description from Eqs.(\ref{propF}-\ref{BC2}), we adapt the method recently introduced for Fabry-Perot cavities (FP)\,\cite{ConfortiFast2025} to ring resonators.
The procedure can be summarized in four steps: (i) perform a change of variables for enforcing spatial periodic boundary conditions to properly define the cavity modes; (ii) simplify the propagation equations for the transformed fields through the good‑cavity approximation; (iii) find the coupled equations for the cavity modes; and (iv) derive a partial differential equation ruling the mean-fields in each direction. These are defined as the sum of the cavity modes with slowly time-varying amplitudes.\\

The appropriate change of variables turns out to be \cite{lugiato_prati_brambilla_2015}:
\begin{align}
  \label{T1}f(z,t)&=e^{\frac{z}{L}\left(\ln \mathcal{R}-i\delta\right)}F(z,t)+\mathcal{T} \frac{z}{L}E_{\rm in},\\
  \label{T2}b(z,t)&=e^{\frac{L-z}{L}\left(\ln \mathcal{R}-i\delta\right)}B(z,t)+\mathcal{T} \frac{L-z}{L}E_{\rm in}^B.
\end{align}
From Eqs. (\ref{BC1},\ref{BC2}), we obtain periodic boundary conditions for the new variables:
\begin{align}
  f(0,t)&=f(L,t),\label{BC1new}\\
  b(0,t)&=b(L,t).\label{BC2new}
\end{align}

We now restrict our analysis to good cavities by assuming $\mathcal{R}\rightarrow 1$, $\mathcal{T}\rightarrow 0$, $\delta\rightarrow 0$.
At zero-th order we have $F\approx f$ and $B\approx b$.
From Eqs.~(\ref{T1}-\ref{T2}), we compute $\partial_z F$, $\partial_t F$, $\partial_z B$, $\partial_t B$ as a function of $f,b$ and their derivatives.
We truncate the resulting expressions at first order in $\mathcal{T}$, $(1-\mathcal{R})$ and $\delta$ and insert them into Eqs.~(\ref{propF},\ref{propB}).
By considering that dispersion and nonlinearity are first-order corrections, we may use zero-order expansions in the dispersive and nonlinear terms. These approximations greatly simplify the propagation equations, which then read:
\begin{align}
  \frac{\partial f}{\partial z}-\frac{\mathcal{T}}{L}E_{\rm in}-\left(\frac{\ln\mathcal{R}}{L}-i\frac{\delta}{L}\right)f+\beta_1\frac{\partial f}{\partial t} \label{propfMF}
  + i\frac{\beta_2}{2}\frac{\partial^2 f}{\partial t^2}&=i\gamma \left(|f|^2+X|b|^2\right)f+\frac{g_b}{2A_{\rm eff}}Qb,\\
  -\frac{\partial b}{\partial z}-\frac{\mathcal{T}}{L}E_{\rm in}^B-\left(\frac{\ln\mathcal{R}}{L}-i\frac{\delta}{L}\right)b+\beta_1\frac{\partial b}{\partial t} \label{propbMF}
  + i\frac{\beta_2}{2}\frac{\partial^2 b}{\partial t^2}&=i\gamma \left(|b|^2+X|f|^2\right)b-\frac{g_b}{2A_{\rm eff}}Q^*f.
\end{align}
%
The modes of the lossless, empty and undriven cavity [$\beta_2=\delta=\gamma=E_{\rm in}=E_{\rm in}^B=g_b=0$, $\mathcal{R}=1$]  can be written as:
\begin{align}
  \label{modeF}f_m(z,t) & =a_m e^{-i\omega_m t }e^{i k_m z }=a_me^{-i\omega_m (t - \beta_1 z)},\\
  \label{modeB}b_m(z,t) & =b_m e^{-i\omega_m t }e^{-i k_m z }=b_me^{-i\omega_m(t + i \beta_1 z)},
\end{align}
where the frequency and wavenumber are:
\begin{equation}
  \omega_m=\frac{2\pi}{\beta_1L}m,\;\;k_m=\beta_1\omega_m=\frac{2\pi}{L}m.
\end{equation}
We have two families of modes with independent amplitudes for the two propagation directions. This is the main difference from FP resonators, where the amplitudes of forward and backward modes are identical\,\cite{Ziani2024}.
We thus separately write the fields in Eqs. (\ref{propfMF},\ref{propbMF}) as the sum of the modes, allowing for a temporal variation of the modal amplitudes:
\begin{align}
  \label{expanF}f(z,t)=\sum_m a_m(t)e^{-i\omega_m (t - \beta_1 z)},\\
  \label{expanb}b(z,t)=\sum_m b_m(t)e^{-i\omega_m(t +  \beta_1 z)}.
\end{align}
To obtain the evolution of the amplitudes $a_m(t)$,
we first apply the expansions Eqs. (\ref{expanF},\ref{expanb}) in Eq. (\ref{propfMF}). The result is then projected onto each forward mode by multiplying the equation by $\frac{1}{L}\exp[i\omega_n (t - \beta_1 z)]$, and integrating from $z=0$ to $z=L$.  After some algebra, we obtain:
\begin{align}\label{modal1}
  \beta_1 \dot a_n -\left(\frac{\ln\mathcal{R}}{L} - i\frac{\delta}{L}\right)a_n+i\frac{\beta_2}{2}&\left(\ddot a_n -2i\omega_n\dot a_n -\omega_n^2a_n\right)-\frac{\mathcal{T}}{L}E_{\rm in}\delta_{n0}=i\gamma \mathcal{K}_n^f+ \frac{g_b}{2A_{\rm eff}} \mathcal{B}_n^f,
\end{align}
where, $\delta_{n0}$ is the Kroneker delta.
Similarly, the evolution of $b_m(t)$ is obtained from Eq.(\ref{propbMF}), through the projection onto the backward modes (multiplication by $\frac{1}{L}\exp[i\omega_n (t + \beta_1 z)]$):
\begin{align}\label{modal1b}
  \beta_1 \dot b_n -\left(\frac{\ln\mathcal{R}}{L} - i\frac{\delta}{L}\right)b_n+i\frac{\beta_2}{2}&\left(\ddot b_n -2i\omega_n\dot b_n -\omega_n^2b_n\right)-\frac{\mathcal{T}}{L}E_{\rm in}^B\delta_{n0}=i\gamma \mathcal{K}_n^b-\frac{g_b}{2A_{\rm eff}} \mathcal{B}_n^b.
\end{align}
In these equations, $\mathcal{K}_n^{f/b}$ and $\mathcal{B}_n^{f/b}$ are the nonlinear forward or backward Kerr and Brillouin coefficients. They are defined as follow.

\begin{itemize}
  \item  \textbf{Kerr term}

The Kerr terms for the forward and backward fields are:

    \begin{align}
      \mathcal{K}_n^f&=\frac{1}{L}\int_{0}^L(|f|^2+X|b|^2)fe^{i\omega_n(t-\beta_1z)}dz=\sum_{n',n''}a_{n'}a_{n''}^*a_{n-n'+n''}+X\sum_{n',n''}b_{n'}b_{n''}^*a_{n+n'-n''}e^{-i2(\omega_{n'}-\omega_{n''})t},\\
      \mathcal{K}_n^b & =\frac{1}{L}\int_{0}^L(|b|^2+X|f|^2)be^{i\omega_n(t+\beta_1z)}dz=\sum_{n',n''}b_{n'}b_{n''}^*b_{n-n'+n''}+X\sum_{n',n''}a_{n'}a_{n''}^*b_{n+n'-n''}e^{-i2(\omega_{n'}-\omega_{n''})t}.
    \end{align}

  \item \textbf{Brillouin term}

The strategy is to find an expression for the acoustic wave $Q$ as a function of the modal amplitudes of the optical fields. To this end, we first decompose the acoustic wave by expanding it over the cavity modes, while keeping the temporal dependence general:
    \begin{equation}
      \label{expanQ}Q(z,t)=\sum_{m}Q_m(t)e^{i k_m z}.
    \end{equation}
We then insert Eq. (\ref{expanQ}) in Eq. (\ref{Q}), and expand the nonlinear term over the modes. Separating the terms oscillating with the same wavenumber we obtain:
    \begin{equation}\label{Qm}
      \left(\frac{\partial^2}{\partial t^2}+\Gamma_b\frac{\partial}{\partial t}+\Omega_b^2\right)Q_m(t)=i\Gamma_b\Omega_b\sum_n a_nb_{m-n}^*e^{-i(2\omega_n-\omega_m)t}.
    \end{equation}
Equation (\ref{Qm}) is a set of forced harmonic oscillators, whose steady-state solutions can be written as
    \begin{equation}
      \label{expanQm}Q_m(t)=\sum_{p}q_{mp}(t)e^{-i \Omega_pt}.
    \end{equation}
We substitute Eq.(\ref{expanQm}) into Eq.(\ref{Qm}) and assume that $q_{mp}(t)$ varies slowly with respect to $\Omega_p$.
By collecting all terms with identical oscillation frequencies, we get $\Omega_p=2\omega_n-\omega_m$ and
    \begin{equation}
      q_{m,2n-m}=iH_b(2\omega_n-\omega_m)a_nb_{m-n}^*,
    \end{equation}
where we have defined the Brillouin response in the frequency domain as:
    \begin{equation}
      H_b(\omega_n)=\frac{\Omega_b\Gamma_b}{\Omega_b^2-\omega_n^2-i\omega_n\Gamma_b}.
    \end{equation}
The acoustic wave envelope has thus the following expression:
    \begin{equation}
      Q(z,t)=i\sum_{n,m}a_nb_{m-n}^*H_b(2\omega_n-\omega_m)e^{-i(2\omega_n-\omega_m)t}e^{i k_m z}.
    \end{equation}

Finally, as for the Kerr terms, the forward Brillouin term $\propto Qb$ in Eq. (\ref{propfMF}) is first expanded over the cavity modes and then projected onto each forward mode. It follows the Brillouin coefficients:
    \begin{align}\label{NBf}
      \mathcal{B}_n^f=\frac{1}{L}\int_{0}^LQbe^{i\omega_n(t-\beta_1z)}dz=i\sum_{n',n''}a_{n'}b_{n''}b^*_{n-n'+n''}H_b(2\omega_{n'}-\omega_{n''}-\omega_n)e^{-i(2\omega_{n'}-\omega_n)t}.
    \end{align}

Similarly, for the backward field, we have:
    \begin{align}\label{NBb}
      \mathcal{B}_{n}^b=\frac{1}{L}\int_{0}^LQ^*fe^{i\omega_n(t+\beta_1z)}dz=-i\sum_{n',n''}a_{n'}^*a_{n''}b_{n-n'+n''}H_b^*(2\omega_{n'}-\omega_{n''}-\omega_{n})e^{2i(\omega_{n'}-\omega_{n''})t}.
    \end{align}
\end{itemize}
We note that since $H_b(\omega)=H_b^*(-\omega)$, it is straightforward to show, via an appropriate change of indices, that $\mathcal{B}_n^{b}(a_{n'},b_{n''})=-\mathcal{B}^f_{n}(b_{n'},a_{n''})$.\\


To obtain the coupled equations for the cavity modes, we insert the nonlinear terms Eqs.(\ref{NBf},\ref{NBb}) in Eqs.(\ref{modal1},\ref{modal1b}). We then
make use of the assumption that the modal amplitudes change slowly over a round-trip (i.e. $|\dot a_n|\ll|\omega_n a_n|$) to simplify the dispersive contribution and to average out the fast oscillations in the nonlinear terms. It follows:
\begin{align}
  \label{modal_fin} \dot a_n +\left(\kappa + i\frac{\delta}{\beta_1L}-i\frac{\beta_2}{2\beta_1}\omega_n^2\right)a_n-\frac{\mathcal{T}}{\beta_1L}E_{\rm in}\delta_{n0}&=i\frac{\gamma}{\beta_1}\left(\sum_{n',n''}a_{n'}a_{n''}^*a_{n-n'+n''}+Xa_n\sum_{n'}|b_{n'}|^2\right)\\
  \nonumber&+i\frac{g_b}{2A_{\rm eff}\beta_1}a_n\sum_{n'}|b_{n'}|^2H_b(\omega_{n}-\omega_{n'}),\\
  \label{modal_fin_b}\dot b_n +\left(\kappa + i\frac{\delta}{\beta_1L}-i\frac{\beta_2}{2\beta_1}\omega_n^2\right)b_n-\frac{\mathcal{T}}{\beta_1L}E_{\rm in}^B\delta_{n0}&=i\frac{\gamma}{\beta_1}\left(\sum_{n',n''}b_{n'}b_{n''}^*b_{n-n'+n''}+Xb_n\sum_{n'}|a_{n'}|^2\right)\\
  \nonumber &+i\frac{g_b}{2A_{\rm eff}\beta_1}b_n\sum_{n'}|a_{n'}|^2H_b(\omega_{n}-\omega_{n'}),
\end{align}%
where we have defined the decay rate $\kappa=-\frac{\ln\mathcal{R}}{\beta_1L}$.
Equations (\ref{modal_fin},\ref{modal_fin_b})  describe the temporal evolution of the slowly varying amplitudes of forward and backward modes. They constitute the coupled-mode description commonly used for micro-resonators.
It is easy to give a physical interpretation of the Brillouin term in these equations. The complex gain experienced by the mode $a_n$ (respectively $b_n$) is given by the discrete convolution of the  power spectrum of the counter-propagating wave $|b_m|^2$ (respectively $|a_n|^2$) and the Brillouin response $H_b(\omega_m)$.\\

As a final step, we derive the equivalent model for slowly varying envelopes of the forward and backward fields \cite{Chembo2013}, defined as follows:
\begin{align}
  \psi_F(z,t)=\sum_m a_m(t) e^{ik_m z},\ \;\;
  \psi_B(z,t)=\sum_m b_m(t)  e^{-ik_m z}.
\end{align}
These fields are periodic with period $L$ ($k_m=2\pi m/L=\beta_1\omega_m$), and the modal amplitudes can be retrieved as follows:
\begin{align}
  \label{Fcoef} a_m(t)=\frac{1}{L}\int_{0}^{L}\psi_F(z,t)e^{-ik_m z}dz, \;\;
  b_m(t)=\frac{1}{L}\int_{0}^{L}\psi_B(z,t)e^{ik_m z}dz.
\end{align}

Multiplying Eq.(\ref{modal_fin}) by $e^{i\beta_1\omega_m z}$ and Eq.(\ref{modal_fin_b}) by $e^{-i\beta_1\omega_m z}$, summing over $m$, and using
\begin{align*}
  \frac{\partial \psi_{F}}{\partial t}=\sum_m \dot a_m(t) e^{ i\beta_1\omega_m z},\ \;\;
  \frac{\partial^n \psi_{F}}{\partial z^n}=\sum_m ( i\beta_1 \omega_m)^n a_m(t)e^{ i\beta_1\omega_m z},
\end{align*}
as well as the analogous relations for $\partial_t \psi_B$ and $\partial_z^n \psi_B$, Eq.(1) of the main manuscript is finally obtained:
%
\begin{align}
  \nonumber T_r\frac{\partial \psi_{F/B}}{\partial t}=-\left(\frac{\Lambda}{2}+i\delta\right)\psi_{F/B}-i\frac{\beta_2L}{2\beta_1^2}\frac{\partial^2 \psi_{F/B}}{\partial z^2}+\sqrt{\theta P_{\rm in}^{F/B}}&+i\gamma L\left(|\psi_{F/B}|^2+2\langle|\psi_{B/F}|^2\rangle\right)\psi_{F/B}
  \\&+i\frac{g_b}{2A_{\rm eff}}\psi_{F/B}\circledast\left[\tilde h_b(\mp z)\cdot\left(\psi_{B/F}^*\circledast \psi_{B/F}(-z)\right)\right],
\end{align}
where $\langle.\rangle=\frac{1}{2}\int_{0}^L(.)dz$
denotes spatial average, $T_r=\beta_1 L$ and $\frac{\Lambda}{2}=\kappa T_r\approx 1-\mathcal{R}$, $\mathcal{T} E_{\rm in}^{F,B}=\sqrt{\theta P_{\rm in}^{F/B}}$.
In this equation, we have defined the (periodic) convolution as
\begin{equation}
  [\psi\circledast \tilde h_b](z)=\int_{0}^L\psi(\xi,t)\tilde h_b(z-\xi)d\xi,
\end{equation}
and the  Brillouin response in the \emph{spatial} domain as:
\begin{equation}
  \label{hB} \tilde h_b(z)=\sum_n\frac{H_b(\omega_n)}{L}e^{-ik_nz}.
\end{equation}
Equation (\ref{hB}) describes the Brillouin reponse in the spatial domain as an $L$-periodic function of $z$, where the Fourier coefficients are  $h_n=H_b(\omega_n)/L$. From the properties of Fourier series, we note that $\tilde h_b(z)=\sum_nh(z-nL)$ is the periodic replication of the inverse Fourier transform in $z$ of the Brillouin response (thus written as a function of the {wavenumber):
  \begin{equation}
    H_b(k/\beta_1)=\frac{\Omega_b\Gamma_b}{\Omega_b^2-(k/\beta_1)^2-i \Gamma_b k/\beta_1}=\int_{-\infty}^{\infty}h(z)e^{ikz}dz
  \end{equation}
  \begin{equation}
    h(z)=\frac{\beta_1\Omega_b\Gamma_b}{\sqrt{\Omega_b^2-(\Gamma_b/2)^2}}e^{-\frac{\Gamma_b}{2}\beta_1 z}\sin\left(\sqrt{\Omega_b^2-(\Gamma_b/2)^2}\,\beta_1z\right)u(z),
  \end{equation}
where $u(z)$ is the Heaviside step function.\\

\newpage

  \section{III. CSs equilibrium positions and impact of a vacancy}

  
  We derive how a vacancy in the soliton lattice induces a distortion in each subsequent CS. 
  To this end, we begin by introducing the interaction potential generated by a single CS, which is of the form (motivated by the experimental fit and numerical computation):
  \begin{equation}
    d(\tau) = \sin(a \tau + \Delta \phi) \exp(-c \tau)  u(\tau).
  \end{equation}
  Here $\tau = -\beta_1 z$ denotes the cavity fast time, $a$ the oscillation frequency of the interaction  ($\approx 2\Omega_b $) and $c \approx \Gamma_b/2 $ its damping rate. $u(\tau)$ is the Heaviside function which ensures causality.     
  In what follows, we neglect the periodic boundary condition, assuming that the cavity round-trip time is much longer than the lifetime of the acoustic wave.
  
  For a single CS, the potential creates a lattice with period $p = 2\pi / a$, within which subsequent solitons can be trapped. Each trapped CS, in turn, contributes to the total interaction potential, so that any given soliton experiences the cumulative effect of all preceding CSs. 
  We consider the total interaction potential in the vicinity of the $n$-th lattice site, denoted $V_n(\tau)$ and valid for $|\tau - n p| \lesssim p$. Owing to the common oscillation frequency of all contributions, we assume that this cumulative potential has the same functional form:
  \begin{equation}
    V_n(\tau) = \sum_{i| \tau_i < np}^\mathrm{CS} d(\tau - \tau_i) = A_n \sin(a \tau + \phi_n) \exp(-c(\tau - n p)).
  \end{equation}

The trapping position of a possible CS is 
\begin{equation}
    \tau_n = n p -\phi_n/a
\end{equation} 
where we neglect the weak correction arising from the decaying exponential envelope.
The presence or absence of a vacancy in the lattice will affect the position of subsequent CSs in distinct ways.\\

\textbf{Occupied n$^\mathrm{th}$ site}\\

If a CS is present, its individual contribution to the potential will modify the potential around the next lattice point $n+1$:
  \begin{align}
    V_{n+1} &= V_n + d (\tau - \tau_n) \notag \\
    &=  \exp(-c (\tau - (n+1) p)) \times \left[
      A_n \sin(a \tau + \phi_n) \exp(-c p)
      + \sin(a \tau + \Delta \phi + \phi_n) \exp(
        -c \Delta_n)
      \right] \notag \\
      &=  \exp(-c (\tau - (n+1) p)) A_{n+1} \sin(a \tau + \phi_{n+1}).
    \end{align}
For convenience, we introduce $\Delta_n = (n+1) p - \tau_n$ and $\xi_n = \exp(-c  (\Delta_n - p) ) / A_n \approx 1/A_n$, 
where the latter approximation holds for $\Delta_n \approx p$ and sufficiently large $A_n$, such that $\xi_n \ll 1$, 
so that the additional contribution of a single soliton constitutes a small perturbation.

Using trigonometric addition formulas, we obtain
\begin{equation}
      A_{n+1} =  A_{n} \exp(-c p) \sqrt{
        (1 +  \xi_n \cos\Delta\phi)^2
        + (\xi_n \sin\Delta\phi)^2
      }
    \end{equation}
    \begin{align}
      \phi_{n+1} =& \tan^{-1}
      \cfrac{\sin(\phi_n) + \xi_n \sin(\Delta \phi + \phi_n) }
      {\cos(\phi_n) + \xi_n \cos(\Delta \phi + \phi_n)} \notag \\
      &= \phi_{n} + \xi_n \sin \Delta \phi + \mathcal{O}(\xi_n^2) \,
    \end{align}

To first order in $\xi_n$, the phase therefore undergoes a constant increment proportional to $\sin(\Delta\phi)$. Since the trapping position depends on $\phi_n$, the position of the subsequent CS is shifted by:

\begin{equation}
    \delta \tau_{n+1} = - \cfrac{\xi_n \sin \Delta\phi}{a}.
\end{equation}

\newpage
\textbf{Fully occupied lattice}\\

We now consider a fully occupied lattice. Translational invariance requires $A_{n+1} = A_{n}$, which yields the self-consistency condition
\begin{align}
  \xi_n = \xi &= -\cos(\Delta \phi) + \sqrt{\cos(\Delta \phi)^2 + \exp(2c p) - 1} \notag\\
  &\approx -\cos(\Delta \phi) + \sqrt{\cos(\Delta \phi)^2 + 2 c p},
\end{align}
where we used $\exp(2c p) \approx 1 + 2c p$ for $c p \approx \pi \Gamma_b / (2\Omega_b) \ll 1$.
Since $\xi_n$ is now constant, the phase increment becomes
\begin{equation}
  \phi_{n+1} - \phi_{n} = \xi \sin(\Delta \phi)
\end{equation}
corresponding to a uniform renormalization of the lattice spacing. The effective period is therefore
\begin{equation}
    p' = \cfrac{p}{ 1 + \xi \sin(\Delta \phi) / 2 \pi}
\end{equation}

\textbf{Unoccupied $n^\mathrm{th}$ site}\\

If a vacancy occurs now at position $n$, the additional contribution $d(\tau-\tau_n)$ is absent. Consequently,
\begin{align*}
       V_{n+1} &= V_n  \\     \phi_{n+1} &= \phi_n \\ A_{n+1} &= \exp(-cp) A_n.
\end{align*}
Compared to the fully occupied case, this corresponds to a missing phase shift $\xi \sin(\Delta\phi)$. All subsequent lattice points are therefore shifted by:
\begin{equation}
    \delta \tau_{n+1} = \cfrac{\xi \sin \Delta\phi}{a}.
\end{equation}
A vacancy thus acts as a localized defect whose effect propagates to all following soliton positions, thereby mediating a long-range coupling between lattice occupancy and positional distortion.\\

\textbf{Finite lattice Segment}\\

For a fully occupied lattice segment surrounded by vacancies, the inter‑soliton delay changes monotonically as it evolves from $p$ to $p'$. This smooth variation reflects the gradual buildup of the phase increment $\phi_{n+1}-\phi_n$ from zero (empty lattice) to its asymptotic constant value in the bulk lattice.
This variation is reflected in the phase of the time oscillation of CSs due to the $S_0$-$S_2$ beat signal. Specifically, one expects a pronounced curvature at the onset of the fully occupied segment, assyptotically evolving to a constant slope. This explains the pattern seen in Fig.\,4(b) of the main paper for -10\,ns\,$<\tau<$\,-5\,ns.

Finally, we note that the periodic boundary conditions of the cavity may introduce a small collective drift of the entire soliton ensemble, as the total accumulated phase must remain compatible with cavity periodicity. Although this global constraint was included when fitting the interaction potential, it does not alter the local vacancy-induced distortion mechanism described above.

\newpage


\section{IV. Brillouin and CS thresholds}

We first compare the input power threshold of the first Brillouin order $S_1$ ($P_b^{th,1}$) with the minimum power ($P_\mathrm{CS}$) to drive a CS. The Brillouin lasing threshold depends on the intra-cavity power and thus on the cavity detuning $\delta$. Assuming that the Brillouin gain peaks at the frequency of a cavity mode, it is given by:
    \begin{equation}
      \cfrac{g_b}{A_\mathrm{eff}} L \cfrac{\theta P_b^{th,1}}{
        \Lambda^2/4 + 4(1 - \Lambda/2) \sin(\delta /2)^2
      } = \Lambda.
      \label{eq_airy}
    \end{equation}
For CSs, the minimum driving power at a given detuning is determined by the saddle‑node bifurcation ($\mathrm{\textit{SN}}_\mathrm{CS}$) that marks the end of the soliton existence range. Without Brillouin interaction it is given by: 

    \begin{equation}
      \delta < \cfrac{\pi^2}{2}  \cfrac{\theta P_\mathrm{CS} \gamma L}{\Lambda^2}.
    \end{equation}

\noindent The CS and Brillouin thresholds are strongly dependent on the cavity loss $\Lambda$ as shown in Figs.\,\ref{fig-sup2:threshold-cs-brillouin}(a-c), where the three possible regimes are identified: Brillouin lasing alone, Brillouin-less cavity solitons, and the coexistence of Brillouin and CSs.
Because the intra-cavity power reaches a minimum at $\delta=\pi$, there exists a loss threshold above which CSs cannot form without Brillouin lasing. Conversely, in low‑loss resonators, this regime becomes accessible at large detunings and low driving powers.
As seen in Fig.\,\ref{fig-sup2:threshold-cs-brillouin}(b), we operated the cavity in the coexistence region.\\   

It is also instructive to examine the driving power required to reach the successive thresholds of the higher‑order Stokes waves
$S_{j}\, (j>1)$. 
Above each threshold, the oscillation condition imposes $P_{Sj-1} - P_{Sj+1}  = \Lambda A_\mathrm{eff}/ (g_b L)$.
As a consequence, at large detunings, as in the experiments, the Brillouin orders emerge in pairs,
[$S_{2n+1}$ and $S_{2n+2}$, see\,Fig.\,\ref{fig-sup2:threshold-cs-brillouin}(d)], with a small hysteresis. This analysis shows that, in the absence of any influence of the CSs on the Brillouin thresholds, four thresholds should be crossed in the experiments. Numerical simulations without CSs confirm this expectation [see Fig.\,\ref{fig-sup3}(e)].
However, the experiments only exhibit the first two Stokes waves [see Figs.\,3\,(b-c) of the main manuscript]. 
Simulations based on the mean‑field model reveal that this discrepancy arises from the large number of CSs circulating in the cavity [Fig.\,\ref{fig-sup3}(d)]. The CSs modify the intracavity spectral power at the driven frequency ($S_0$), thereby increasing the input power required to reach the Brillouin thresholds.  
    
    \begin{figure}[h]
      \centering
      \includegraphics[width=14cm]{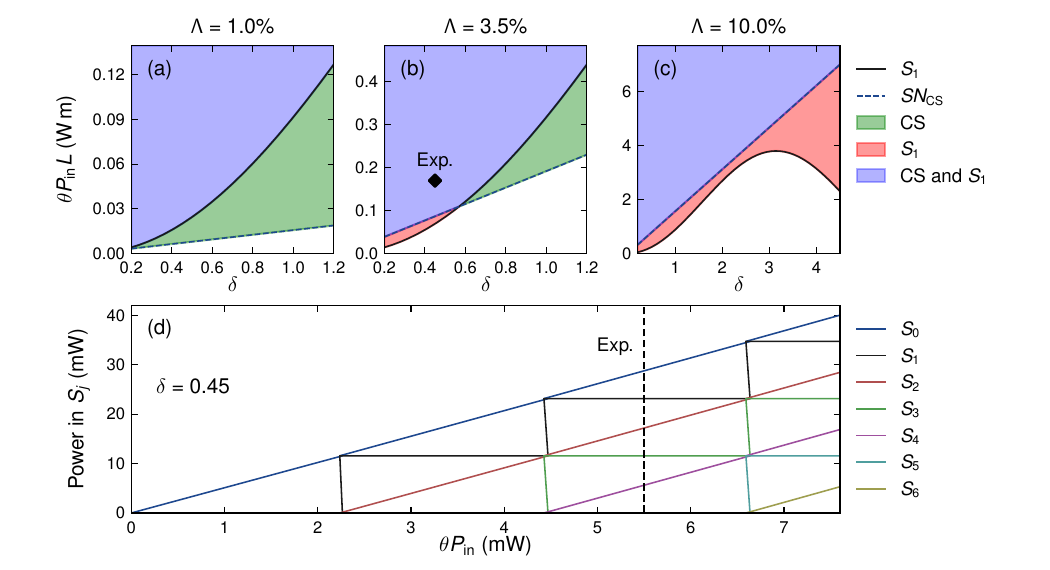}
      \caption{
        (a)-(c) First-order Brillouin threshold ($S_1$), $P_\mathrm{in} = P_b^{th,1}$, and CS threshold ($\mathrm{\textit{SN}}_\mathrm{CS}$), $P_\mathrm{in} = P_\mathrm{CS}$, for different cavity loss $\Lambda$. The shaded areas distinguish the diffent regimes: CS, only cavity solitons, $S_1$ Brillouin lasing only, CS and $S_1$ coexistence of soliton with Brillouin lasing.   
        (d) Input pump power thresholds for the formation of the $j$-th Stokes wave $S_j$ in the resonator at $\delta=0.45$ and $\Lambda = 3.5\%$. The diamond in (c) and the vertical line in (d) represent the experimental conditions. 
      }
      \label{fig-sup2:threshold-cs-brillouin}
    \end{figure}

\newpage
    
\section{V. Simulation parameters and simulation method}

\textit{Simulation parameters---}The simulation parameters are given in Table \ref{tab:sim_params}.  
We selected the set of parameters that provides the best agreement between the numerical and experimental results. This explains the small discrepancies with the experimental values for the input coupler and driving power ($\theta=1.3\%$ from the datasheet and $P_\mathrm{in}=600$\,mW in the experiment).
We note that this choice of simulation parameters is not unique, as several nearby combinations of $\gamma$, $\theta$, $\alpha$, and $P_{\mathrm{in}}$ can reproduce the same observed dynamics.
We further note that the actual values of the fiber and component parameters are subject to uncertainties, and that parasitic effects, such as time‑varying polarization‑dependent loss (PDL), might slightly modify the cavity loss experienced by the driven modes ($\Lambda$) between the characterization stage and the actual experiments.
 
\noindent In simulations, a small amount of white noise $P_\mathrm{noise}$ was added to the cavity field at each roundtrip to mimic the residual drive ASE floor and to seed the Brillouin lasing. The value of $P_\mathrm{noise}$ is somewhat arbitrary and does not significantly affect the simulation outcomes.\\

\textit{Simulation method---}A 5(6) adaptive Runge-Kutta method in the Interaction Picture (RKIP) \cite{hultFourthOrderRungeKutta2007} was used to numerically solve the coupled equations. We stress that due to the counter-propagating nature of the fields, the simulation window must encompass the entire cavity round-trip time ($\approx 150$\,ns). This requirement is computationally expensive. To address this challenge, the inherent parallelism of these equations was leveraged to implement the simulation on a GPU, resulting in a speed-up factor of $10$ to $100$, depending on the hardware. For example, simulating 4000 roundtrips requires approximately $\qty{3}{\min}$ on an NVIDIA RTX 4080 for $2 \times 2^{19}$ simulation points.\\

\vspace{0.5cm}

    \begin{table}[h]
      \centering
      \begin{tabular}{l c l}
        \toprule
        Parameter & {Value} & Unit \\
        \midrule
        $\beta_{1}$      & \qty{4.87e-9}{}   & \si{\second\per\meter} \\
        $\beta_{2}$      & \qty{-2.1e-26}{}  & \si{\second\squared\per\meter} \\
        $\gamma$         & \qty{1.3e-3}{}  & \si{\per\watt\per\meter} \\
        \midrule
        $g_b /A_\mathrm{eff}$ & 0.10       & \si{\per\watt\per\meter} \\
        $\nu_b = \Omega_{b} / 2\pi$ & 10.85 & \si{\giga \hertz} \\
        $\Gamma_{b} / 2\pi$ & 28.0 & \si{\mega \hertz} \\
        \midrule
        $L$              & 30.7      & \si{\meter} \\
        $\mathcal{F}$    & 177       & -- \\
        $\alpha= \Lambda/2 = \pi/\mathcal{F} $ & \qty{1.8e-2}{} & -- \\
        $\theta$         & \qty{1.0e-2}{}  & -- \\
        $P_{\mathrm{in}}$& 550    & \si{\milli \watt} \\
        $P_{\mathrm{noise}}$ & 100 & \si{\micro \watt} \\

        \bottomrule
      \end{tabular}
      \caption{Simulation parameters.}
      \label{tab:sim_params}
    \end{table}


\section{VI. Additional figure}

The additional Fig.\,\ref{fig-sup3} provides a further comparison between the experiments and simulations based on the mean-field model Eqs.(A1)-(A2) of the main manuscript, as well as insight into the pattern formation. Fig.\,\ref{fig-sup3}(a) shows the powers, experimentally measured in the cavity through port (PD$_2$, forward signal, blue curve) and in the third port of the first circulator (PD$_1$, backward signal, red curve), as the detuning is scanned from left to rigth. 
This is the same scan as in Fig.\,2\,(b) of the main manuscript. The corresponding simulation with the parameters of Table\,\ref{tab:sim_params} is given in Fig.\,\ref{fig-sup3}(b). As can be seen, the agreement is very good, with the simulation reproducing faithfully the transient starting dynamics of the Brillouin laser during the detuning scan, or the soliton step and the decay of the backward Brillouin signal for positive detunings. This provides a direct validation of the mean‑field model presented in this study.

From simulations, we can analyze in the spectral domain the interplay between the Brillouin cascade and the dynamics of the CSs pattern formation during a resonance scan. Fig.\,\ref{fig-sup3}(c) shows that the $S_2$ threshold is crossed before the spontaneous modulation instability (MI) can take place in the resonator [and that it is simultaneously crossed  with the $S_1$ threshold by comparison with Fig.\,\ref{fig-sup3}(b)]. Further increasing the detuning, the thresholds of $S4$ and $S_6$ are crossed and four-wave mixing generates the anti-Stokes waves. As can be seen these spectral lines seed the soliton spectral comb thus setting the pattern periodicity close to twice the Brillouin frequency shift. The spectral density at $\delta=0.45$ is display in Fig.\,\ref{fig-sup3}(d) for the forward and the backward intra-cavity fields. We see that only the first two Brillouin thresholds ($S_1$, $S_2$) are crossed at this detuning when CSs are circulating in the cavity. In contrast, up to four Stokes waves are generated without cavity solitons. This shows the competition between CSs and the Brillouin cascade lasing.  

    \begin{figure}[h]
      \centering
      \includegraphics[width=15cm]{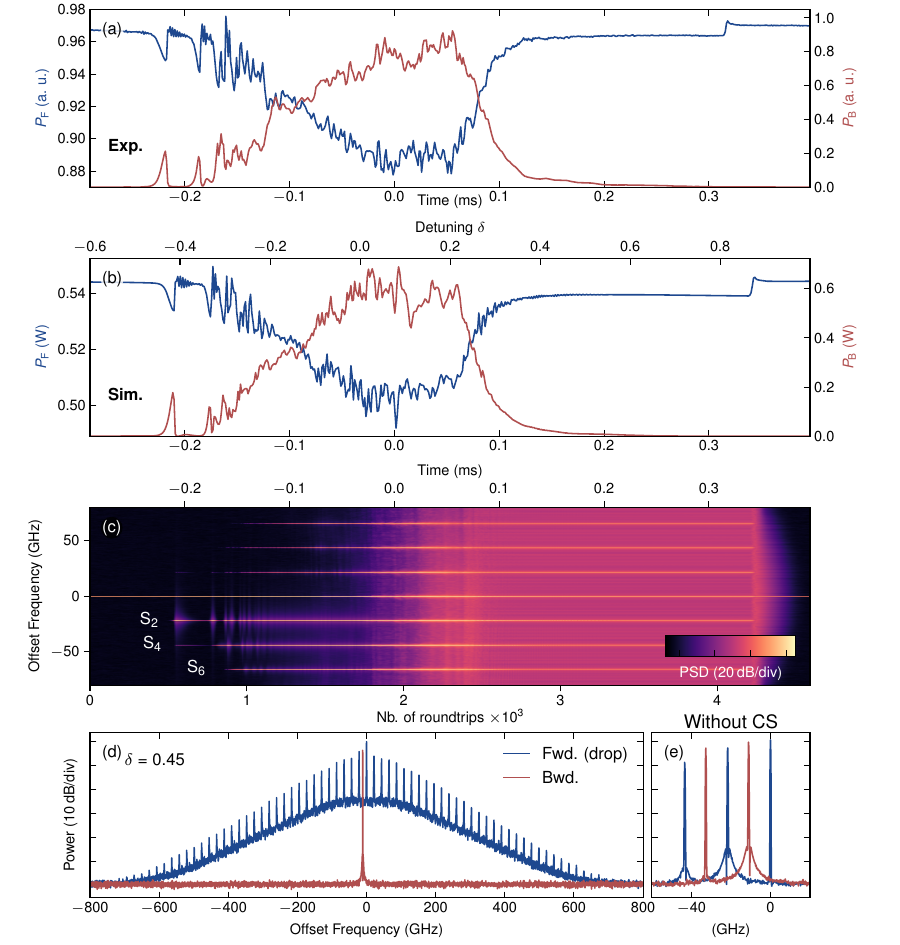}
      \caption{
        (a) Experimentally recorded transmitted ($P_F$) and reflected ($P_B$)
        signals when scanning the frequency resonance.
        (b) Corresponding numerical simulation with the mean-field model.
        (c) Evolution of the cavity power spectral density of $\psi_F$ during the scan shown in (b).
        (d) Simulated PSDs in a stabilized cavity at $\delta= 0.45$ after a detuning \textit{scan-stop}.
        (e) Same as in (d) but without Kerr effect ($\gamma=0$), thus without CSs. 
      }
      \label{fig-sup3}
    \end{figure}

    \bibliography{supplemental_ref, referencesV2}